\documentclass[preprint,superscriptaddress]{revtex4-2} 
\usepackage{graphicx} % Required for inserting images
\usepackage{algorithm}
\usepackage{algpseudocode}
\usepackage[margin=1in]{geometry}
\usepackage{amssymb}
\usepackage{amsmath}
\usepackage{mathrsfs}
\usepackage{placeins}
\usepackage{caption3}
\usepackage{wrapfig}
\usepackage{caption} 
\usepackage{subcaption}
\usepackage{xcolor}
\usepackage{contour}
\usepackage{color}
\usepackage{outlines}
\usepackage[bookmarksnumbered=true,breaklinks=true,colorlinks,citecolor=blue,linkcolor=blue]{hyperref}
\usepackage{lipsum}

\begin{document}
\title{Hierarchical equivariant graph neural networks for forecasting collective motion in vortex clusters and microswimmers}
\author{Alec J. Linot}
\affiliation{Mechanical and Aerospace Engineering, University of California, Los Angeles, CA 90095, USA}
\author{Haotian Hang}
\affiliation{Aerospace and Mechanical Engineering, University of Southern California, Los Angeles, CA 90089, USA}
\author{Eva Kanso}
\affiliation{Aerospace and Mechanical Engineering, University of Southern California, Los Angeles, CA 90089, USA}
\author{Kunihiko Taira}
\affiliation{Mechanical and Aerospace Engineering, University of California, Los Angeles, CA 90095, USA}

\begin{abstract}
Data-driven modeling of collective dynamics is a challenging problem because emergent phenomena in multi-agent systems are often shaped by long-range interactions among individuals. For example, in bird flocks and fish schools, long-range vision and flow coupling drive individual behaviors across the collective.
Such collective motion can be modeled using graph neural networks (GNNs), but GNNs struggle when graphs become large and often fail to capture long-range interactions. 
Here, we construct hierarchical and equivariant GNNs, and show that these GNNs accurately predict local and global behavior in systems with collective motion. As representative examples, we apply this approach to simulations of clusters of point vortices and populations of microswimmers. For the point vortices, we define a local graph of vortices within a cluster and a global graph of interactions between clusters. 
For the microswimmers, we define a local graph around each microswimmer and a global graph that groups long-range interactions. We then combine this hierarchy of graphs with an approach that enforces equivariance to rotations and translations. This combination results in a significant improvement over a fully-connected GNN. For point vortices, our method conserves the Hamiltonian over long times, and, for microswimmers, our method predicts the transition from aggregation to swirling.
\end{abstract}
\maketitle

A wide range of non-living and living systems exhibit collective motion in which emergent behavior appears from simple interaction laws \citep{Vicsek2012}. Examples include shaken metallic rods \citep{Kudrolli2010}, micromotors \citep{Ibele2009}, bacteria colonies \citep{Wu2000}, birds flocking \citep{Ballerini2008}, and pedestrian dynamics \citep{Moussad2011}. 
Often, the interaction laws dictating these systems are unknown, making them an ideal candidate for data-driven modeling. However, although many data-driven methods exist for forecasting dynamics, including recurrent neural networks \citep{Hochreiter}, neural ODEs \citep{Chen2018}, and reservoir computing \citep{Lukovsevivcius}, which have proven effective in chaotic systems \citep{Vlachas2019,Pathak2018,Linot2020,Linot2021}, turbulent fluid flows \citep{Borrelli2022,Linot2023}, and weather forecasting \citep{Maulik2020,Arcomano2022}, these methods are not suited for learning the interaction laws that lead to emergent collective phenomena in multi-agent systems. Here, to overcome these challenges, we propose the use of graph neural networks (GNN) that account for the hierarchical and equivariant nature of many collective dynamics systems.

GNNs are popular in modeling multi-agent systems and simulations on unstructured grids because the number of agents in the graph can vary and the spatial relationship between agents can be incorporated into the graph. For example, GNNs have been used in particle-based physics simulations \citep{Sanchez-Gonzalez2020,Bhattoo2023} of Lagrangian fluid simulations \citep{Li2022} and granular flows \citep{Choi2024}. \citet{Xiong2023} extended these ideas to vortex datasets by using a detection network to identify vortices and then forecasting the detected vortices using a GNN similar to \citet{Battaglia2016}. Another use of GNNs for fluid simulations is for mesh-based models \citep{Pfaff2021}, by treating each grid point as a node in a graph. \citet{Peng2022} applied GNNs for reduced-order modeling of convex channel flow and expansion flows.

Standard GNNs are purely data-driven and must learn physical properties from that data, which presents a challenge if these properties must be exactly satisfied. 
One such property in many physical systems is symmetry. 
For example, the dynamics of fluid in a straight pipe are equivariant to rotations and reflections and invariant to translations down the pipe \citep{Willis2013}. Exactly satisfying these properties in data-driven tools presents a major challenge. Many GNN methods present specific architectures to guarantee types of equivariance such as E(3) equivariance \citep{Thomas2018,Fuchs2020}, E(n) equivariance \citep{Santorras2021}, steerable E(3)-equivariant \citep{Brandstetter2022,Toshev2023}, and SE(3)-equivariance \citep{Xu2024}. These types of methods have also been combined with other popular methods like neural ODEs \citep{Liu2024} and transformers \citep{Liao2023,Liao2024} for multi-agent problems. Furthermore, equivariant GNNs have been applied to irregular meshes for fluid simulations \citep{Lino2022,Shankar2023}.

Most of these methods account for equivariance by using specific choices of the GNN architecture. Alternatively, equivariance can be enforced by mapping the data to an invariant space and mapping the resulting output back. This has the advantage of enforcing that \emph{any} GNN structure guarantees equivariance. We will perform this mapping by modifying a method based on principal component analysis (PCA) \citep{Xiao2019}, to address the sign ambiguity problem discussed in \citep{Li2021}.

Although the GNN approaches have been used in many problems, less effort has been put toward data-driven modeling of collective motion problems where long-range order appears. \citet{Wang2022} used GNNs to predict an order parameter in the Vicsek model \citep{Vicsek1995} but did not perform time forecasting. \citet{Heras2019} used deep attention networks and videos of zebrafish to predict the likelihood of a zebrafish turning right after some time. They further used these results to infer interaction rules between fish. For predicting the time-evolution of collective dynamics problems,  \citet{Ha2021} developed a method called AgentNet, similar to graph attention networks \citep{velickovic2018graph}, which they predicted the dynamics of cellular automata, the Vicsek model, active Ornstein-Uhlenbeck particles, and bird flocking. 

In what follows, we show the importance of using a hierarchy of local and global graphs and of enforcing equivariance when forecasting collective motion problems that exhibit long-range order.  Importantly, our framework is agnostic to the details of the GNN, making it widely applicable. We test this method on two problems that show the breadth of applicability of our approach: clusters of point vortices and microswimmers. The first represents an example of a large-scale conservative system, the second represents an example of an out-of-equilibrium, non-conservative, active system. In the first one, interactions decay as $1/r$, in the second, interactions decay as $1/r^2$. The latter exhibits a bifurcation as a function of the control parameter. First, we outline our approach in Sec.\ \ref{sec:GNN} and \ref{sec:Equiv}. We then apply the method to clusters of vortices in Sec.\ \ref{sec:Vortex} and to a model of microswimmers in Sec.\ \ref{sec:Swimmer}. We conclude with a discussion of these results in Sec.\ \ref{sec:Discussion}, and outline the detailed methods in Sec.\ \ref{sec:Methods}.

\section{Results}
\label{sec:Results}

\subsection{Graph Neural Networks for Collective Motion}
\label{sec:GNN}

Many systems that exhibit collective motion can be modeled with a system of ordinary differential equations (ODE)
\begin{equation} \label{eq:ODE}
    \dfrac{d\mathbf{r}_i}{dt}=f_i(\mathbf{r}_1,\dots,\mathbf{r}_N),
\end{equation}
where $\mathbf{r}_i(t)\in\mathbb{R}^d$ are the properties of agent $i$ of $N$ agents. 
As mentioned in the introduction, Eq.\ \ref{eq:ODE} may not always be known. Thus, we aim to approximate Eq.\ \ref{eq:ODE} directly from data. In particular, we aim to approximate Eq.\ \ref{eq:ODE} by training a method that predicts the right-hand-side of the ODE
\begin{equation} \label{eq:GNNODE}
    \dfrac{d\mathbf{\tilde{r}}_i}{dt}=\tilde{f}_i(\mathbf{r}_1,\dots,\mathbf{r}_N;\theta),
\end{equation}
where $\tilde{\cdot}$ indicates an approximation, to minimize the loss given by
\begin{equation} \label{eq:Loss}
    \mathcal{L}=\dfrac{1}{MNd}\sum_{j=1}^M \sum_{i=1}^N \left|\left| \left. \dfrac{d\mathbf{r}_{i}}{dt}\right|_{t_j} -\left. \dfrac{d\mathbf{\tilde{r}}_{i}}{dt}\right|_{t_j} \right|\right|^2.
\end{equation}
In Eq.\ \ref{eq:GNNODE}, $\theta$ corresponds to the training parameters of the data-driven model that are updated to minimize the loss over a batch of $M$ data points. Here, we construct Eq.\ \ref{eq:GNNODE} using graph neural networks.
GNNs are a class of data-driven methods that apply neural networks to graph data. A graph $G=(V,E)$ consists of a set of nodes (vertices) $V=\{v_1,v_2,\dots,v_N\}$ connected by edges $E=\{e_{1,1},e_{1,2},\dots,e_{N,N}\}$ \citep{Diestel2017}. On this graph, we define node weights (i.e., signals) $\mathbf{n}_i: V\rightarrow \mathbb{R}^d$ and edge weights $W_{i,j}: E\rightarrow \mathbb{R}$. These edge weights are also known as the weighted adjacency matrix. They are used to construct the normalized graph Laplacian $L=I-D^{-1/2}WD^{-1/2}$, where $D_{i,i}=\sum_j W_{i,j}$ is the diagonal degree matrix and $I$ is the identity. Here we consider undirected graphs, which results in a symmetric adjacency matrix, allowing us to write the graph Laplacian in this form. The eigenvectors of the normalized graph Laplacian are known as graph Fourier modes, which enable us to perform graph convolutions \citep{Defferrard2016}.

Our objective is to perform a series of GNN operations to map from a set of initial graphs whose nodal values contain information on properties of the agents $\mathbf{r}_i$ to a final graph whose nodal values output the estimated dynamics $d\mathbf{\tilde{r}}_i/dt$. Similar to deep convolutional neural networks, we perform this mapping through graph convolutions in which we expand the nodal values of the graph from $\mathbf{n}_i\in\mathbb{R}^d$ to graphs with nodal values $\mathbf{h}^{(j)}_i\in\mathbb{R}^{d_j}$. We repeat this process for multiple layers, and end by mapping the $K^\text{th}$ graph $\mathbf{h}^{(K)}_i\in\mathbb{R}^{d_K}$ to the output $\mathbf{o}_i\in\mathbb{R}^{d_o}$ predicts $d\mathbf{\tilde{r}}_i/dt$. In this process, we introduce nonlinearity after each graph convolution operation through an activation function, and we compute edge weights based on the distance between nodes. In Sec.\ \ref{sec:MethodsConv}, we describe these operations in more detail.

\begin{figure}
    \centering
    \includegraphics[width=.45\textwidth]{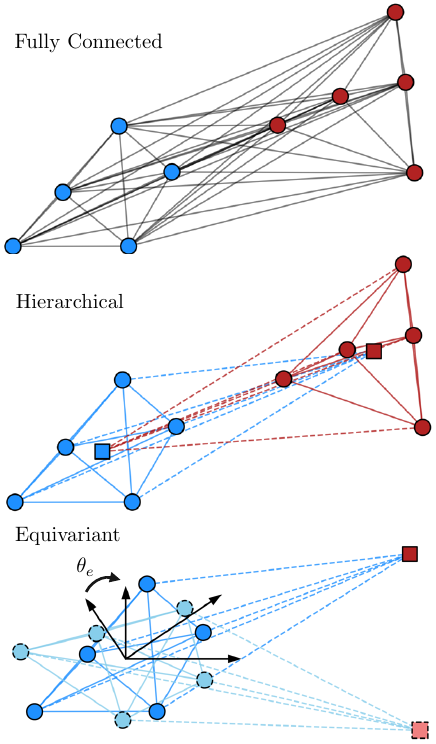}
    \caption{Three styles of graphs. The fully-connected graph connects every node to every other node. The hierarchical graph connects all nearby edges in a local graph (solid lines) and aggregates long-range properties in a global graph (dashed lines). The equivariant graph contains the same connections as either of the two graphs, but maps the nodes to a rotation-invariant reference frame (light colors).}
    \label{fig:Graph}

    \captionsetup[subfigure]{labelformat=empty}
    \begin{subfigure}[b]{0\textwidth}\caption{}\vspace{-10mm}\label{fig:Grapha}\end{subfigure}
    \begin{subfigure}[b]{0\textwidth}\caption{}\vspace{-10mm}\label{fig:Graphb}\end{subfigure}
    \begin{subfigure}[b]{0\textwidth}\caption{}\vspace{-10mm}\label{fig:Graphc}\end{subfigure}
    \vspace{-5mm}
\end{figure}

\begin{figure*}
    \centering
    \includegraphics[width=\textwidth]{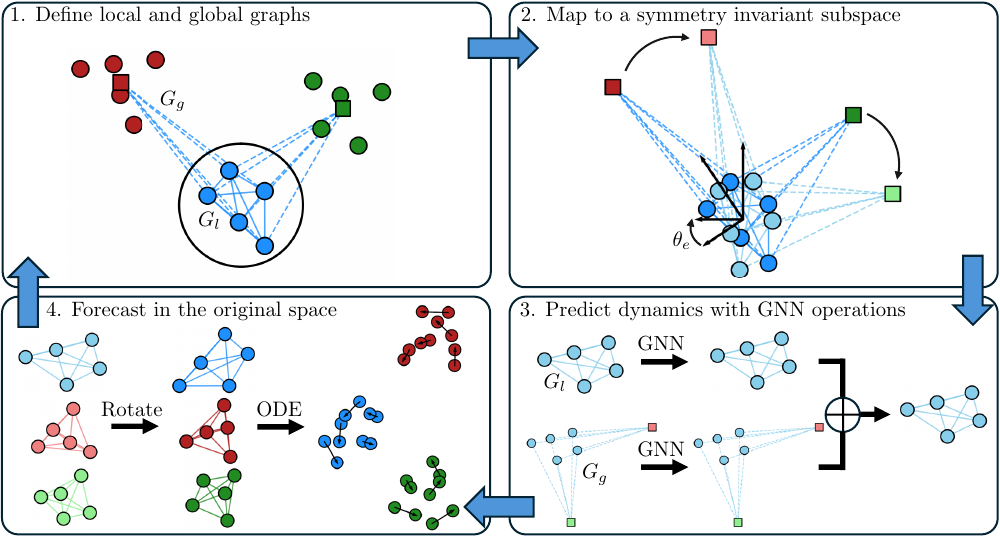}
    \caption{The present hierarchical and equivariant GNN method illustrated using the point vortex problem. Solid lines are local edges and dashed lines are global edges. Dark colors lie in the original space and light colors lie in the invariant subspace. }
    \label{fig:Diagram}

    \captionsetup[subfigure]{labelformat=empty}
    \begin{subfigure}[b]{0\textwidth}\caption{}\vspace{-10mm}\label{fig:Diagrama}\end{subfigure}
    \begin{subfigure}[b]{0\textwidth}\caption{}\vspace{-10mm}\label{fig:Diagramb}\end{subfigure}
    \begin{subfigure}[b]{0\textwidth}\caption{}\vspace{-10mm}\label{fig:Diagramc}\end{subfigure}
    \vspace{-5mm}
\end{figure*}

A key factor in the performance of the GNN is how the graphs are constructed. Figure \ref{fig:Graph} presents three styles of graph construction. One approach is to generate a fully-connected graph with the agents' properties $\mathbf{n}_i=\mathbf{r}_i$.
Unfortunately, this graph construction results in an expensive forward pass of the GNN because the network is dense, and hinders training as the GNN must learn which connections are more or less important. 
Another approach is to generate a graph in which only local connections within some radius $R$ (or k-nearest neighbors) are retained. This approach reduces the computational cost by removing unimportant connections but could omit the importance of many weak long-range interactions. 
Instead of adopting these approaches, we construct a hierarchy of local $G_l$ (short-range interactions) and global $G_g$ (long-range interactions) graphs. 
The advantage of this approach is that it accounts for long-range interactions while decreasing computational costs and improving training performance. While here we only consider constructing a hierarchy of graphs at two scales, the approach easily extends to graphs at more scales by performing the graph convolutions at each scale.

\subsection{Enforcing Equivariance}
\label{sec:Equiv}

Consistent with the physical properties of the collective dynamics, the GNNs should retain invariance to translations and equivariance to rotations. For our GNN to satisfy these properties for some abstract group $g \in \mathcal{G}$, then for some group action $T_g$ on the input, there exists a group action on the output $S_g$ that satisfies 
\begin{equation} \label{eq:GNNODEEquiv}
    S_g\left(\tilde{f}_i([\mathbf{r}_1,\dots,\mathbf{r}_N];\theta)\right)=\tilde{f}_i(T_g([\mathbf{r}_1,\dots,\mathbf{r}_N]);\theta).
\end{equation}
For illustrative purposes, consider the case where the agents' properties correspond to the agents' location in two dimensions (i.e., $\mathbf{r}_i=[x_i,y_i]^T$). In this case, the translation operation on the input is $T_g(\mathbf{r}_i)=[x_i+g_x,y_i+g_y]^T$, and the translation operation on the output $S_g$ is the identity (this is also known as invariance). When all the agents translate by some fixed amount there is no change in the vector field. We can also write out the rotation operation on the input as 
\begin{equation} \label{eq:Rot}
    T_g(\mathbf{r}_i)=R(\theta_g) \mathbf{r}_i=
    \begin{bmatrix}
    \cos (\theta_g)       & -\sin (\theta_g)  \\
    \sin (\theta_g)       & \cos (\theta_g) 
\end{bmatrix}
    \begin{bmatrix}
    x_i  \\
    y_i 
\end{bmatrix},
\end{equation}
which is equivalent to rotating the vector field, instead of the input, with $S_g(\tilde{f}_i(\mathbf{r}_i;\theta))=R(\theta_g)\tilde{f}_i(\mathbf{r}_i;\theta)$.

To enforce equivariance to rotations and invariance to translations, we must either design a GNN that satisfies Eq.\ \ref{eq:GNNODEEquiv} by construction, or we must map the input to an invariant subspace $\mathbf{\hat{r}}_i$, and then apply the appropriate actions to the output. We take the latter approach as it enforces equivariance with \emph{any} GNN structure. This approach is often taken to map the solutions of partial differential equations, with continuous spatial symmetry, to an invariant subspace using the method-of-slices \citep{Froehlich2012,Budanur2015}.   

The objective when defining a translation and rotation invariant state representation $\mathbf{\hat{r}}_i=\mathcal{I}(\mathbf{r}_i)$ is to find a mapping such that $\mathbf{\hat{r}}_i$ does not change for any $T_g$. This can be achieved by defining $\mathbf{\hat{r}}_i=R(-\theta_e)(\mathbf{r}_i-\mathbf{r}_c)$, where $\mathbf{r}_c$ is a point in space that we center the data about, and $\theta_e$ is a unique phase. Both $\mathbf{r}_c$ and $\theta_e$ depend on $\mathbf{r}_i$. Then, if we approximate the dynamics with \begin{equation} \label{eq:RotDyn}
    \dfrac{d\mathbf{\tilde{r}}_i}{dt}=\tilde{f}'_i(\mathbf{r}_i;\theta)=R(\theta_e)\tilde{f}_i(\mathbf{\hat{r}}_i;\theta),
\end{equation} 
we guarantee that $\tilde{f}'_i$ will automatically satisfy Eq.\ \ref{eq:GNNODEEquiv} regardless of the GNN we use for $\tilde{f}_i$. We discuss the specific methods for computing $\mathbf{r}_c$ and $\theta_e$ when describing our test cases.

We summarize the steps for our hierarchical equivariant GNN (HE-GNN) in Fig.\ \ref{fig:Diagram}:
1) construct a hierarchy of local and global graphs ($G_l$ and $G_g$), 2) map the graph to a rotational and translational invariant subspace by centering the data and rotating by $\theta_e$,
3) input the rotational and translational invariant graphs into a GNN, and 4) rotate the output of the GNN back to the original orientation for forecasting. We perform the same GNN operations on the local and global graphs. 
In what follows, we demonstrate the capability of the HE-GNN method to predict the collective dynamics of point vortices and microswimmers. All results shown are on test datasets not used during the GNN training.

\begin{figure*}
    \centering
    \includegraphics[width=\textwidth]{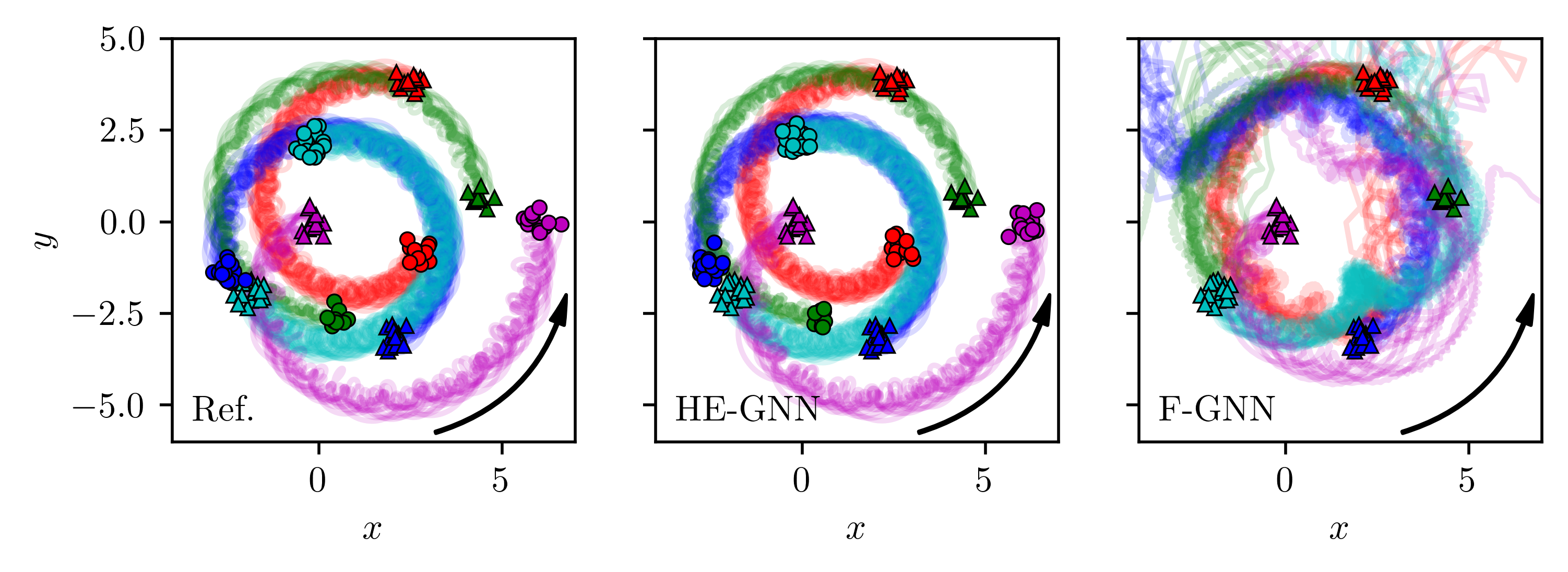}

    \captionsetup[subfigure]{labelformat=empty}
    \begin{picture}(0,0)
    \put(-210,170){\contour{white}{ \textcolor{black}{a)}}}
    \put(-50,170){\contour{white}{ \textcolor{black}{b)}}}
    \put(95,170){\contour{white}{ \textcolor{black}{c)}}}
    \end{picture} 
    \begin{subfigure}[b]{0\textwidth}\caption{}\vspace{-10mm}\label{fig:VortexTraja}\end{subfigure}
    \begin{subfigure}[b]{0\textwidth}\caption{}\vspace{-10mm}\label{fig:VortexTrajb}\end{subfigure}
    \begin{subfigure}[b]{0\textwidth}\caption{}\vspace{-10mm}\label{fig:VortexTrajc}\end{subfigure}
    \vspace{-5mm}

    \caption{Vortex dynamics given by (a) the reference solution, (b) the hierarchical and equivariant GNN, and (c) the fully-connected GNN. The triangles and circles indicate the initial and final positions. Colors differentiate clusters.}
    \label{fig:VortexTraj}
\end{figure*}
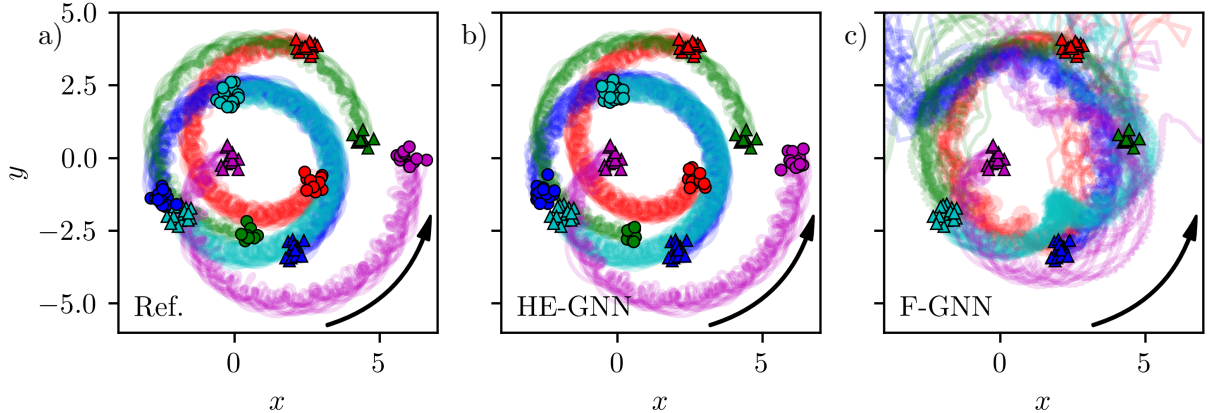

\subsection{Point Vortices}
\label{sec:Vortex}

\begin{figure*}
    \centering
    \includegraphics[width=\textwidth]{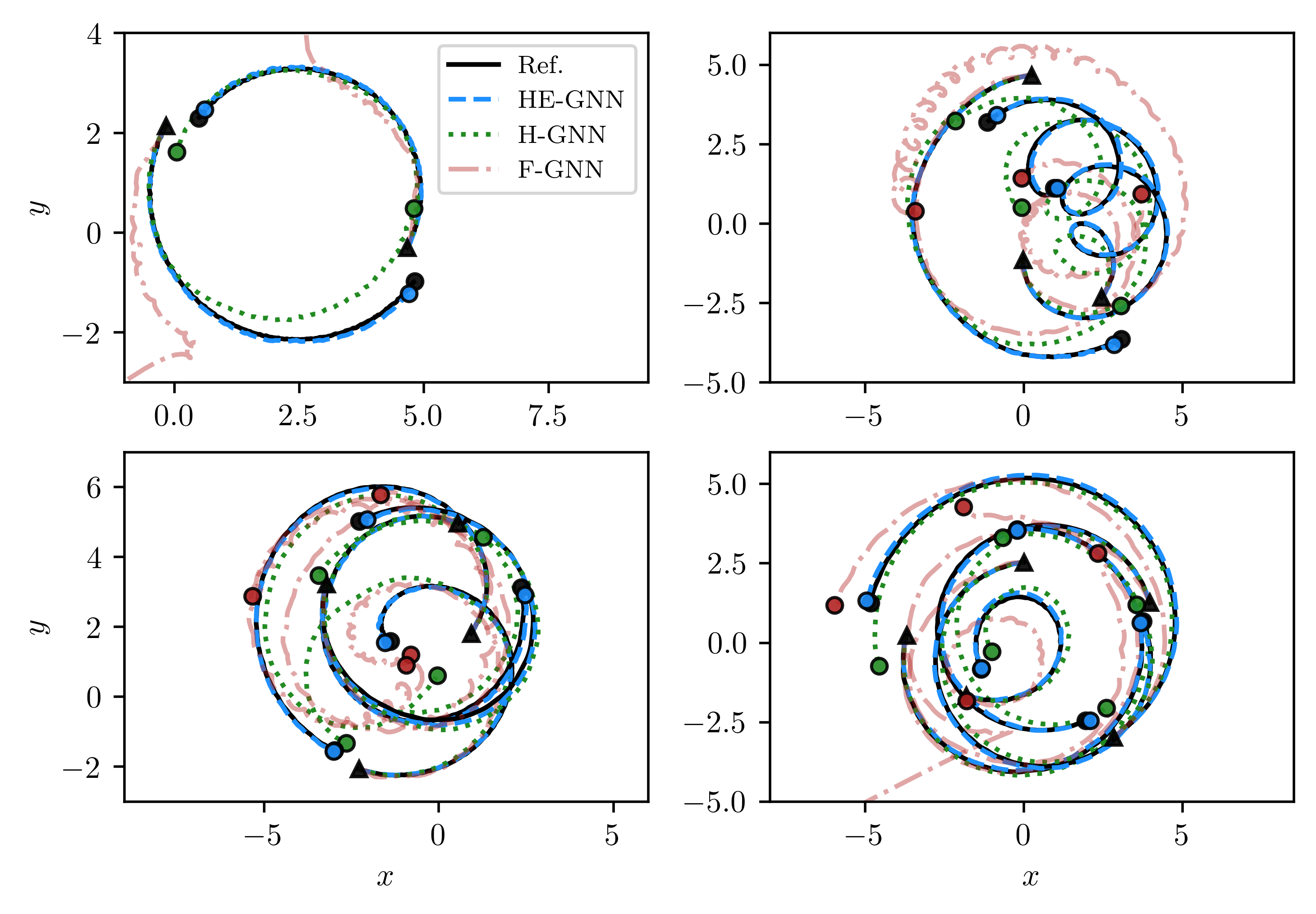}

    \captionsetup[subfigure]{labelformat=empty}
    \begin{picture}(0,0)
    \put(-210,325){\contour{white}{ \textcolor{black}{a)}}}
    \put(10,325){\contour{white}{ \textcolor{black}{b)}}}
    \put(-210,175){\contour{white}{ \textcolor{black}{c)}}}
    \put(10,175){\contour{white}{ \textcolor{black}{d)}}}
    \end{picture} 
    \begin{subfigure}[b]{0\textwidth}\caption{}\vspace{-10mm}\label{fig:VortexCenta}\end{subfigure}
    \begin{subfigure}[b]{0\textwidth}\caption{}\vspace{-10mm}\label{fig:VortexCentb}\end{subfigure}
    \begin{subfigure}[b]{0\textwidth}\caption{}\vspace{-10mm}\label{fig:VortexCentc}\end{subfigure}
    \begin{subfigure}[b]{0\textwidth}\caption{}\vspace{-10mm}\label{fig:VortexCentd}\end{subfigure}
    \vspace{-5mm}

    \caption{Trajectories of centroids for cases with (a) two, (b) three, (c) four, and (d) five clusters from test simulations and GNN models. The triangles and circles indicate the initial and final positions.}
    
    \label{fig:VortexCent}
\end{figure*}

First, we test our model on predicting the dynamics of clusters of point vortices. This is an important application for fluid dynamics problems because many complex flow phenomena can be modeled using vortex methods including separated flows \citep{Cortelezzi1993}, the flow around wind turbines \citep{Chatelain2017}, and free jet flows \citep{Samarbakhsh2019}. A review of these methods can be found in \citep{Eldredge2019,Mimeau2021}. In these flows, large-scale structures can be well-represented by the interactions of many point vortices, thus we test our model on predicting the dynamics of clusters of vortices that are governed by the Biot-Savart law \citep{Saffman_1993}. 
We investigate the interactions of vortex clusters (clouds) by generating two to five clusters each with ten to twenty vortices randomly placed within a cluster for both training and testing. Each vortex has a randomly chosen positive strength (circulation) and position within a cluster. Due to one-sided strength and the initial cluster spacing the clusters stay coherent for long periods of time. We chose these settings to parallel the sparse network methods used in \citep{Nair2015,Meena2018} on these datasets. Figure \ref{fig:VortexTraja} shows example trajectories of this system with five clusters.

We predict the evolution of each vortex cluster using the same GNN. The nodal values $\mathbf{n}_i$ of the local graph $G_l$ correspond to the vortex position and strength ($\gamma_i$), such that $\mathbf{n}_i=[x_i,y_i,\gamma_i]^T$. In this dataset, we do not vary $\gamma_i$ over time, so the GNN only forecasts the dynamics of the vortex position (i.e., $[d\tilde{x}_i/dt,d\tilde{y}_i/dt]^T=\tilde{f}_i(G_l,G_g;\theta)$). The method can account for $\gamma_i$ that varies in time, but we consider a dataset where this is not the case for consistency with \citep{Nair2015,Meena2018}. 

The nodal values of the global graph $G_g$ are $\mathbf{n}_i=[x_i,y_i,\gamma_i]^T$ inside the local cluster and the average position and sum of strength outside of the local cluster $\mathbf{n}_i=[\bar{x}_i,\bar{y}_i,\sum_{k\in C_i} \gamma_k]^T$, where $C_i$ is the set of indices for one of the clusters outside of the local cluster.
Figure \ref{fig:Diagram} shows the local graph with solid lines and the global graph with dotted lines. In the local graph, we connect all nodes, and, in the global graph, we connect all the local nodes to the global nodes. This is a simple way of defining edges in the graphs, but many other approaches work. For example, if the clusters are large, connections in the local graph could be determined by a radius or k-nearest neighbors. Additionally, we define the adjacency matrix to consist of weights $W_{i,j}=1/||\mathbf{x}_i-\mathbf{x}_j||^2$. We also tested $W_{i,j}=1/||\mathbf{x}_i-\mathbf{x}_j||$, which had around two times greater error. To map to the rotation-translation invariant subspace, we center about the mean $\mathbf{r}_c=(1/N_C)\sum_{i\in C_l}\mathbf{r}_i$, where $N_C$ is the number of points in the local cluster and $C_l$ are the indices of points in the local cluster. We then define the angle $\theta_e$ based on the angle of the principal components of the local cluster similar to \citet{Xiao2019}. We describe how we compute this angle in more detail in Sec.\ \ref{sec:MethodsAngle}.

With these graphs, we use standard neural network optimization procedures to update the weights of the GNN to minimize the loss in Eq.\ \ref{eq:Loss}. All models use the same training data, which consists of multiple time series of two to five clusters of ten to twenty vortices. 
We detail the GNN architecture and dataset in Sec.\ \ref{sec:MethodsVortex}. Following model training, we compute trajectories of the model by evolving an initial condition forward using the GNN ODE with the same ODE solver used in generating test data. We then compare the model trajectories with unseen test data. 

We first validate the model performance by comparing the model trajectories to an example trajectory in Fig.\ \ref{fig:VortexTraj}. The example trajectory appears in Fig.\ \ref{fig:VortexTraja}, the HE-GNN model trajectory appears in Fig.\ \ref{fig:VortexTrajb}, and the fully-connected GNN (F-GNN) trajectory appears in Fig.\ \ref{fig:VortexTrajc}. 
The F-GNN is a single GNN that uses all edge weights between vortices. This graph is centered by the mean of all the centroids but does not account for rotational equivariance. The HE-GNN accurately tracks the centroid locations of the test data, whereas the F-GNN dramatically diverges from the test data. Although, in theory, the fully-connected model should be able to track the test data, the training performance of the fully-connected model is much worse because the hierarchical and equivariant properties of the system must be learned during training, whereas these properties are directly enforced in the HE-GNN model.

Now that we have shown in Fig.\ \ref{fig:VortexTraj} that qualitatively the vortex clusters appear to align with one another, we next show how accurate this centroid tracking is as we vary the GNN model and the number of clusters. We examine the centroid tracking because we are interested in the collective dynamics of the vortices, but not necessarily how accurately we track each and every vortex.
Figure \ref{fig:VortexCent} shows the efficacy of different GNN models in tracking the centroid location for two, three, four, and five clusters. The three models we compare include a fully-connected GNN, a hierarchical GNN (H-GNN), and a HE-GNN. 
The H-GNN does not account for rotational equivariance, but is centered, and otherwise matches the setup of the HE-GNN. In Fig.\ \ref{fig:VortexCent}, the hierarchical models do not diverge, whereas the F-GNN diverges with two and five clusters. Furthermore, the hierarchical models show similar error as the number of clusters varies, despite the additional complexity and network size with more clusters. However, Fig.\ \ref{fig:VortexCent} shows a clear improvement in performance when accounting for equivariance, highlighting the advantage of our HE-GNN approach.

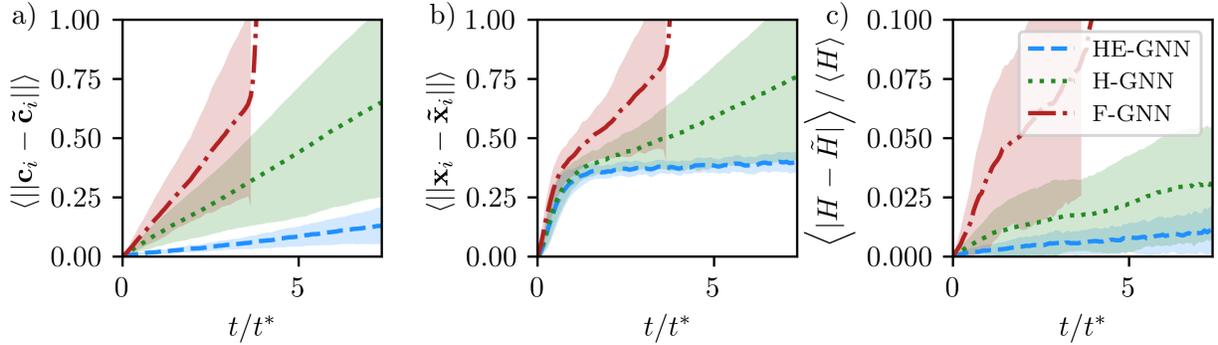
\begin{figure*}
    \centering
    \includegraphics[width=\textwidth]{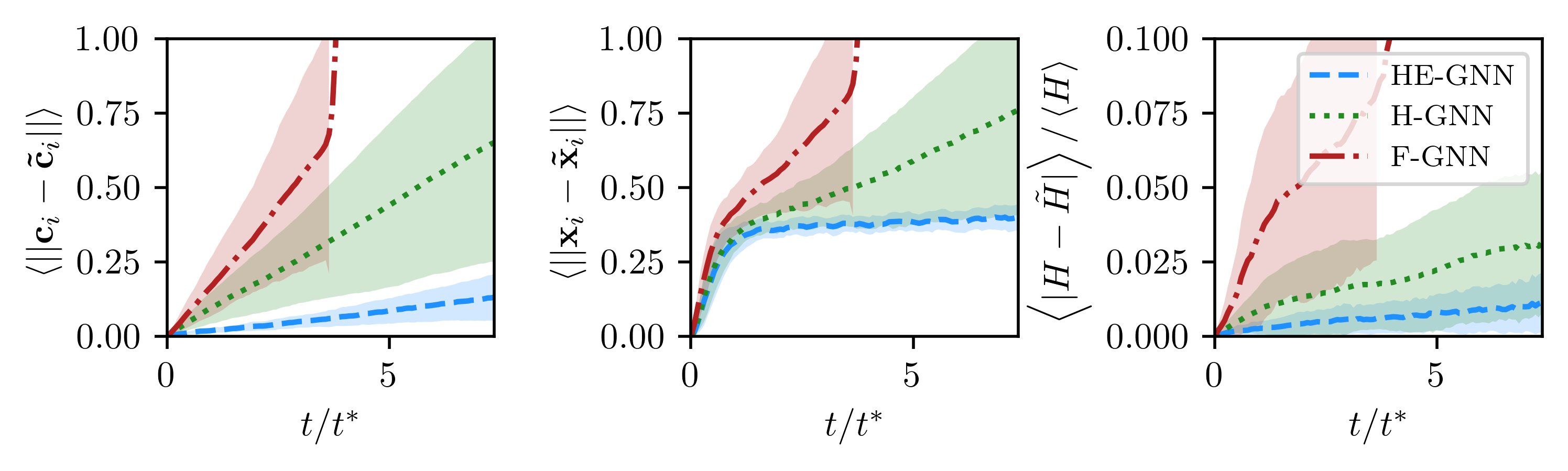}
    \captionsetup[subfigure]{labelformat=empty}
    \begin{picture}(0,0)
    \put(-223,150){\contour{white}{ \textcolor{black}{a)}}}
    \put(-65,150){\contour{white}{ \textcolor{black}{b)}}}
    \put(85,150){\contour{white}{ \textcolor{black}{c)}}}
    \end{picture} 
    \begin{subfigure}[b]{0\textwidth}\caption{}\vspace{-10mm}\label{fig:VortexEnsa}\end{subfigure}
    \begin{subfigure}[b]{0\textwidth}\caption{}\vspace{-10mm}\label{fig:VortexEnsb}\end{subfigure}
    \begin{subfigure}[b]{0\textwidth}\caption{}\vspace{-10mm}\label{fig:VortexEnsc}\end{subfigure}
    \vspace{-5mm}

    \caption{Ensemble averaged tracking error of (a) centroids and (b) inidividual vortices. (c) normalized ensemble averaged error in the Hamiltonian. Shading indicates the standard deviation.}
    
    \label{fig:VortexEnsemble}
\end{figure*}

We further validate the efficacy of the HE-GNN model by considering the tracking performance ensemble averaged over 50 different test initial conditions. These initial conditions vary the cluster count, the centroid locations, the vortex locations, and the vortex strength. To perform this averaging, we need a proper notion of time. When vortex clusters are close together, or there are many vortex clusters, the clusters move faster than when the clusters are far apart, or there are fewer clusters. 
To account for this, we compute the timescale $t^*=D/U_{\text{avg}}$, where $D$ is the cluster diameter and $U_{\text{avg}}$ is the average centroid velocity. This timescale indicates the average time required for a cluster to travel one diameter. 

The ensemble-averaged centroid tracking, vortex tracking, and Hamiltonian conservation are shown in Fig.\ \ref{fig:VortexEnsemble}. The centroid tracking error supports the results shown in Fig.\ \ref{fig:VortexCent}. The HE-GNN substantially outperforms both the F-GNN and the H-GNN. At short times, the models differ in the rate at which the error increases in the three models, and, at long times, the solutions from the F-GNN diverge, while the H-GNN exhibits a larger variance than the HE-GNN. Although the centroid tracking is improved by the HE-GNN (Fig.\ \ref{fig:VortexEnsb}), the tracking of individual agents is similar between all three models. This error plateaus for the HE-GNN and grows for the other two models. These two results highlight that the HE-GNN accurately captures the collective motion of the centroids, but all models perform similarly in the instantaneous tracking of vortices.

Another crucial quantity in correctly capturing the physics of this problem
is the Hamiltonian 
\begin{equation}
    H=\dfrac{1}{4\pi}\sum_{i,j=1,i\neq j}^N \gamma_i\gamma_j\log \left( ||\mathbf{x}_i-\mathbf{x}_j||\right).
\end{equation}
This quantity is conserved and depends on all interactions making it an important test for validating the models. Figure \ref{fig:VortexEnsc} shows the normalized error in the Hamiltonian.
The F-GNN quickly diverges, which also causes the Hamiltonian to diverge because the vortex distances become large. Again, the H-GNN improves upon the F-GNN, and the HE-GNN model shows the best results. The HE-GNN model on average only has $\sim 1\%$ error in the Hamiltonian after moving an averaged distance of over 6 diameters away, which confirms the importance of embedding equivarience in the GNN formulation.

\subsection{Microswimmers}
\label{sec:Swimmer}

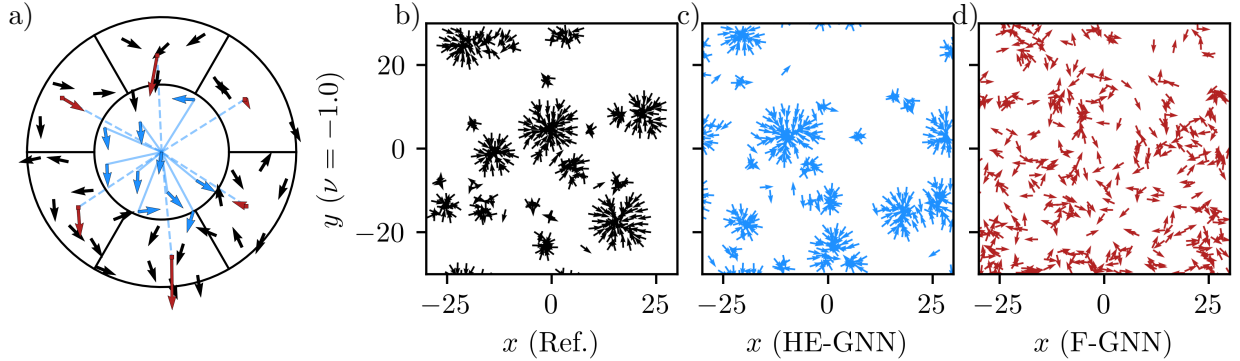
\begin{figure*}
    \centering
    \includegraphics[width=\textwidth]{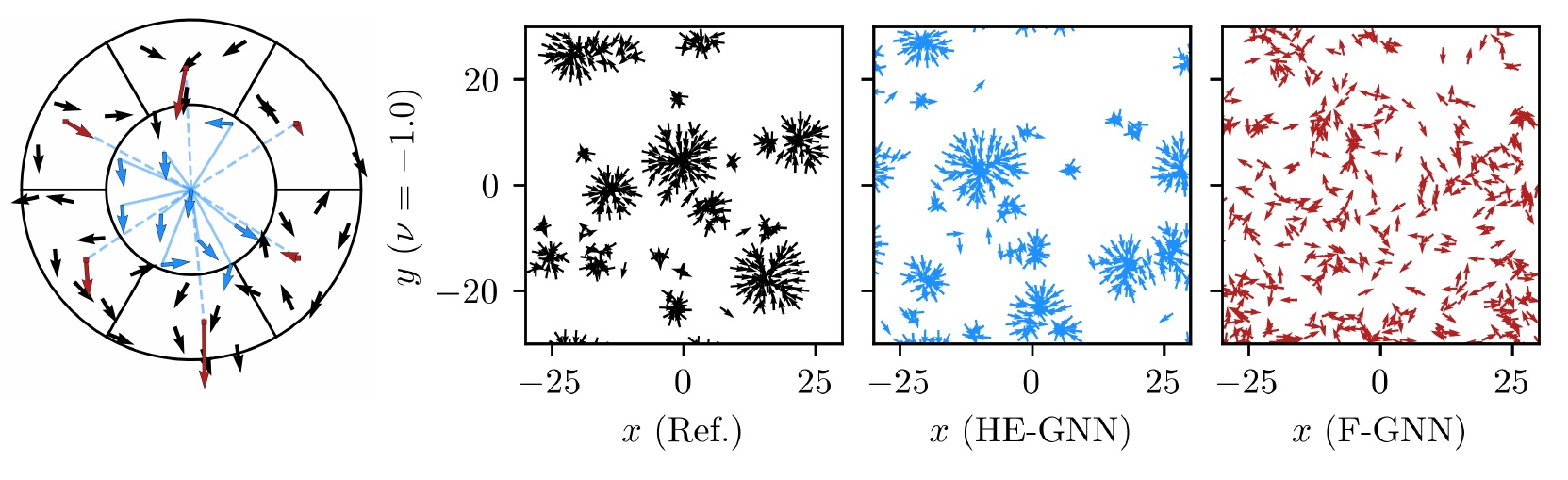}

    \captionsetup[subfigure]{labelformat=empty}
    \begin{picture}(0,0)
    \put(-230,157){\contour{white}{ \textcolor{black}{a)}}}
    \put(-84,157){\contour{white}{ \textcolor{black}{b)}}}
    \put(23,157){\contour{white}{ \textcolor{black}{c)}}}
    \put(127,157){\contour{white}{ \textcolor{black}{d)}}}
    \end{picture} 
    \begin{subfigure}[b]{0\textwidth}\caption{}\vspace{-10mm}\label{fig:FishDiaga}\end{subfigure}
    \begin{subfigure}[b]{0\textwidth}\caption{}\vspace{-10mm}\label{fig:FishDiagb}\end{subfigure}
    \begin{subfigure}[b]{0\textwidth}\caption{}\vspace{-10mm}\label{fig:FishDiagc}\end{subfigure}
    \begin{subfigure}[b]{0\textwidth}\caption{}\vspace{-10mm}\label{fig:FishDiagd}\end{subfigure}
    \vspace{-5mm}
    
    \caption{(a) construction of the local (solid line) and global graphs (dotted line) for microswimmers. (b)-(d) show examples of the long-time dynamics ($t=100$) for the test data at $\nu=-1.0$, the HE-GNN model, and the F-GNN model, respectively.}
    \label{fig:FishDiag}
\end{figure*}

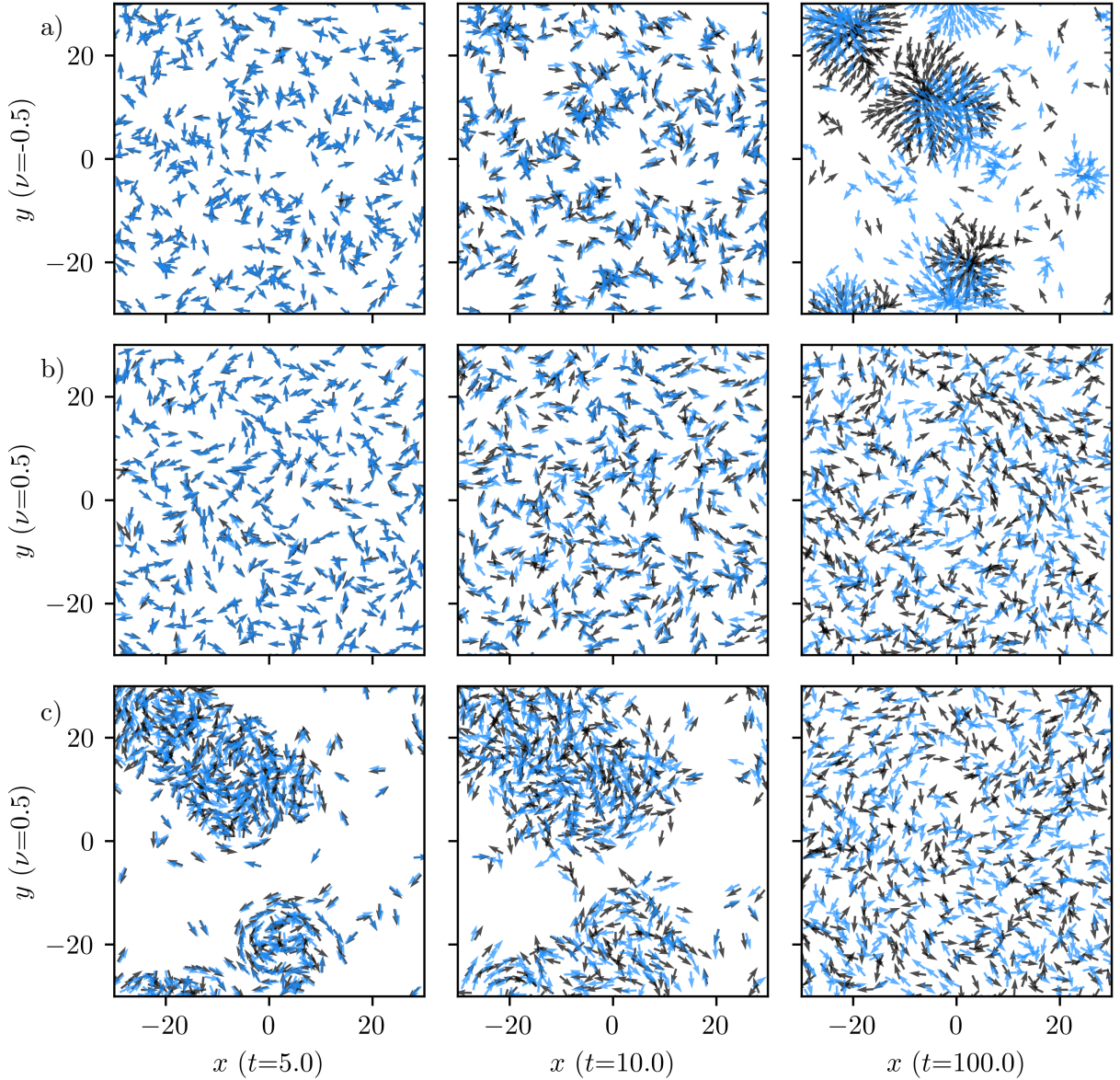
\begin{figure*}
    \centering
    \includegraphics[width=\textwidth]{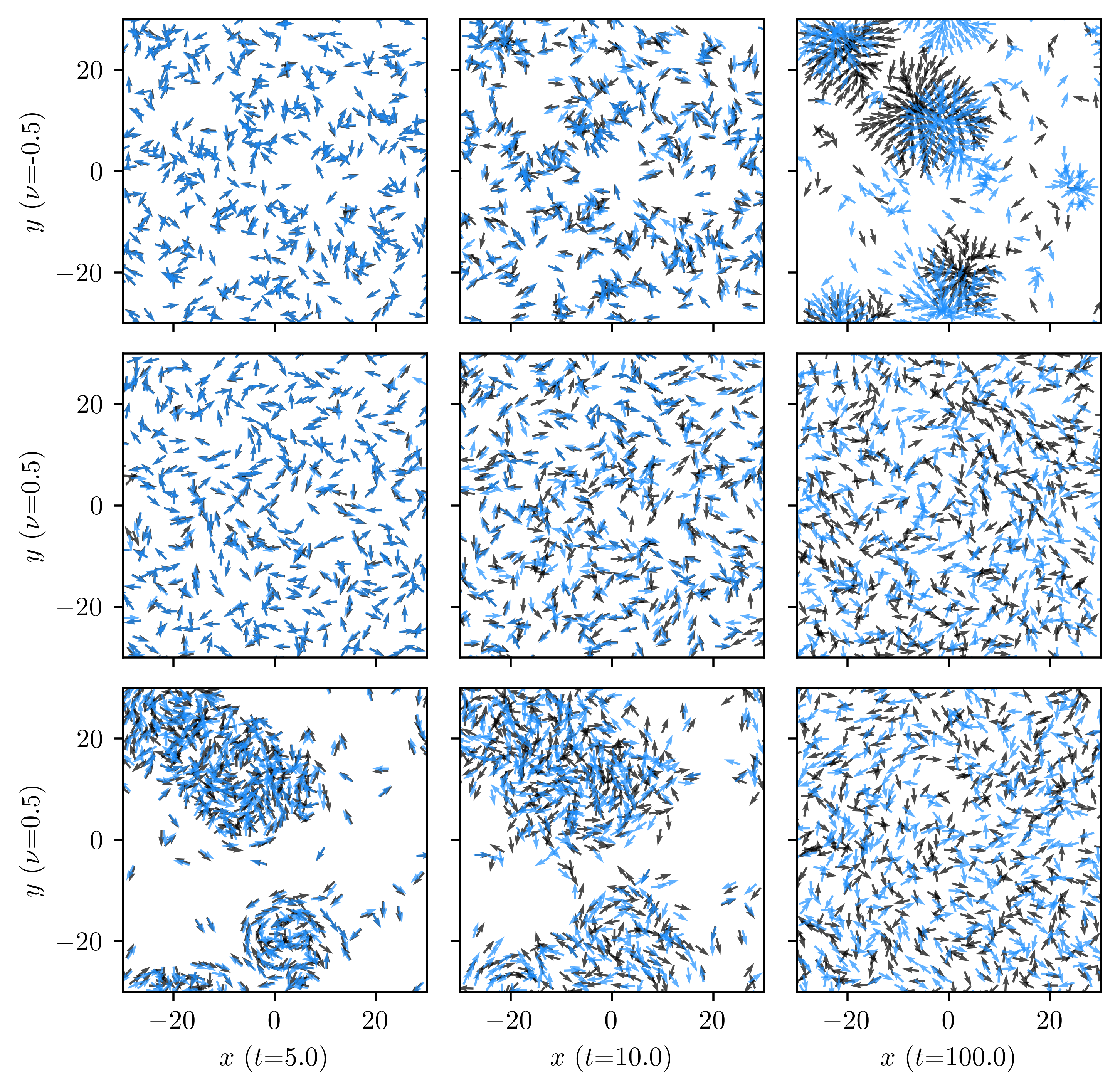}

    \captionsetup[subfigure]{labelformat=empty}
    \begin{picture}(0,0)
    \put(-210,459){\contour{white}{ \textcolor{black}{a)}}}
    \put(-210,318){\contour{white}{ \textcolor{black}{b)}}}
    \put(-210,177){\contour{white}{ \textcolor{black}{c)}}}
    \end{picture} 
    \begin{subfigure}[b]{0\textwidth}\caption{}\vspace{-10mm}\label{fig:Traja}\end{subfigure}
    \begin{subfigure}[b]{0\textwidth}\caption{}\vspace{-10mm}\label{fig:Trajb}\end{subfigure}
    \begin{subfigure}[b]{0\textwidth}\caption{}\vspace{-10mm}\label{fig:Trajc}\end{subfigure}
    \vspace{-5mm}
    
    \caption{Snapshots of microswimmers at various rotational mobility coefficients. Test data is in black and HE-GNN prediction is in blue. (a) and (b) begin from the same random initial condition discussed in Sec.\ \ref{sec:MethodsSwimmer}. (c) is a time series starting from the end condition of (a).}
    \label{fig:Traj}
\end{figure*}

Next, we test the HE-GNN setup on the dynamics of microswimmers in a doubly periodic domain. These microswimmers move due to a combination of self-propulsion and hydrodynamic interactions.
Hydrodynamic interaction plays a dominant role in the collective behaviors of microswimmers and has been considered with Stokesian dynamics~\cite{Brady1988,Elfring2022}, potential flow models~\cite{Oza2019,Heydari2024,Tsang2014,Tsang2016}, and lubrication theory~\cite{Takagi2014}. One common property of these models is the superposition of the flow field generated by each swimmer. Here, we considered a potential dipole model, which we describe in detail in Sec.\ \ref{sec:MethodsSwimmer}. 
Predicting the far-field all-to-all interaction is challenging because there is no inherent graph structure in the model, unlike local behavior models like the Vicsek model or the 3A model~\cite{Vicsek1995,Couzin2003,Filella2018}, where we could restrict our model to only local networks. 
Thus, this microswimmer model allows us to test whether aggregating long-range interactions with our hierarchical approach can capture these far-field effects.

For the microswimmers, our macroscopic property of interest is the phase transition from aggregation to swirling motion as we vary the microswimmers rotational mobility coefficient $\nu$. There is no rotation of microswimmers at $\nu=0$. Above $\nu=0$ the system exhibits swirling, and below $\nu=0$ the system exhibits aggregation \citep{Tsang2014}. 
The microswimmers have clearly defined orientation angles, but do not have a clearly defined hierarchical structure because the swimmers do not always lie within a distinct cluster. We address this issue by defining a local and global graph for every microswimmer.

The local graph consists of connections between the microswimmer whose dynamics we want to predict and all other microswimmers in a radius $R_1$.
We also tested fully connecting all microswimmers the radius $R_1$, which increased computation time and did not improve predictive capabilities. Outside of this region, we aggregate the effect of many microswimmers to create a global graph. 
Here, we define a second region between $R_1$ and $R_2$ and split this region into $S=12$ slices.
Within these slices, we average the microswimmer properties for the global graph. Figure \ref{fig:FishDiaga} shows an example of the connectivity of the local and global graphs. We will show the benefits of incorporating a hierarchical structure into the model. We will not show all of the possible ways one could construct a global graph. However, one alternative could be to perform K-means clustering to identify relevant regions instead of segmenting the domain into predefined regions. We will consider fixing the radii ($R_1$ and $R_2$) first, and then vary them at the end of this section.

The microswimmers have orientations $\alpha_i$ that vary with time. As such, we define agent properties $\mathbf{r}_i=[x_i,y_i,\alpha_i]^T$. We do not include $\nu$ as an agent property because we fix this parameter for each case.
For the local graph $G_l$, the nodal values are $\mathbf{n}_i=[x_i,y_i,\cos(\alpha_i),\sin(\alpha_i)]^T$. We use $\cos(\alpha)$ and $\sin(\alpha)$ so that the GNN output changes continuously. Otherwise, the GNN output would be discontinuous as $\alpha$ varies outside $0$ to $2\pi$. For the global graph $G_g$, we use nodal properties $\mathbf{n}_i=[\bar{x}_i,\bar{y}_i,\overline{\cos(\alpha_i)},\overline{\sin(\alpha_i)},\bar{U}]^T$, where $\bar{\cdot}$ is the average of all agents within one of the $S$ slices and $\bar{U}$ is the magnitude of the vector assuming all agents have a magnitude of 1. We define the adjacency matrix based on the distance between the central agent ($i$) and the other $j$ agents $W_{i,j}=1/||\mathbf{x}_i-\mathbf{x}_j||^2$ for the local graph, and between the central agent and the mean locations $W_{i,j}=1/||\mathbf{x}_i-\bar{\mathbf{x}}_j||^2$ for the global graph. Finally, we enforce equivariance by centering both the local and global graphs about the central ($i$) microswimmer ($\mathbf{r}_c=\mathbf{r}_i$) and rotating the graphs such that this microswimmer has an angle of 0 ($\theta_e=-\alpha_i$). 

As all microswimmers have the same value of $\nu$, 
we incorporate $\nu$ into the GNN by appending it onto $\mathbf{h'}^{(K)}_i=[\nu,\mathbf{h}^{(K)}_i]^T$, and then predicting $d\mathbf{\tilde{r}}_i/dt$ by inputting  $\mathbf{h'}^{(K)}_i$ through a dense neural network. We train the GNN models using a dataset that contains 21 trajectories each with 400 microswimmers. Each trajectory has a different mobility coefficient $\nu$ evenly sampled between $-1$ and $1$. We again use standard neural network optimization procedures to train GNNs (architectures described in Sec.\ \ref{sec:MethodsSwimmer}) to minimize the loss in Eq.\ \ref{eq:Loss}.

First, we investigate the efficacy of the HE-GNN model in comparison to a fully-connected GNN model. In the F-GNN, we connect all swimmers in the periodic box and use their shortest distance across all boundaries for computing the edge weights. We perform the same graph convolutions as in the local portion of the HE-GNN model and append $\nu$ in the same fashion. Figure \ref{fig:FishDiag} compares the performance of the models at long times for strongly aggregating dynamics. The HE-GNN model correctly predicts the strong aggregation behavior, while the F-GNN fails to aggregate.

Next, we test the ability of the HE-GNN model to capture the phase change and the sensitivity of the HE-GNN model to new initial conditions. Figures \ref{fig:Traja} and \ref{fig:Trajb} compare time series of microswimmers from the test data and the HE-GNN model at $\nu=-0.5$ and $\nu=0.5$, respectively. This HE-GNN model uses $R_1=20$ and $R_2=30$. At short times, the model closely tracks the location and orientation of the microswimmers, and, at long times, the model captures the aggregation ($\nu=-0.5$) and swirling ($\nu=0.5$) of the microswimmers. Furthermore, the HE-GNN model is insensitive to the initial condition. Figure \ref{fig:Trajc} shows trajectories starting from the final condition of Fig.\ \ref{fig:Traja} with $\nu=0.5$. Here the aggregation structures begin to disassemble and transition towards swirling dynamics. This type of dynamics does \emph{not} exist in the training dataset.
Despite this, the model accurately captures the rate at which microswimmers expand out of the clusters at short times, and the swirling motion at long times. 

\begin{figure}
    \centering
    \includegraphics[width=.5\textwidth]{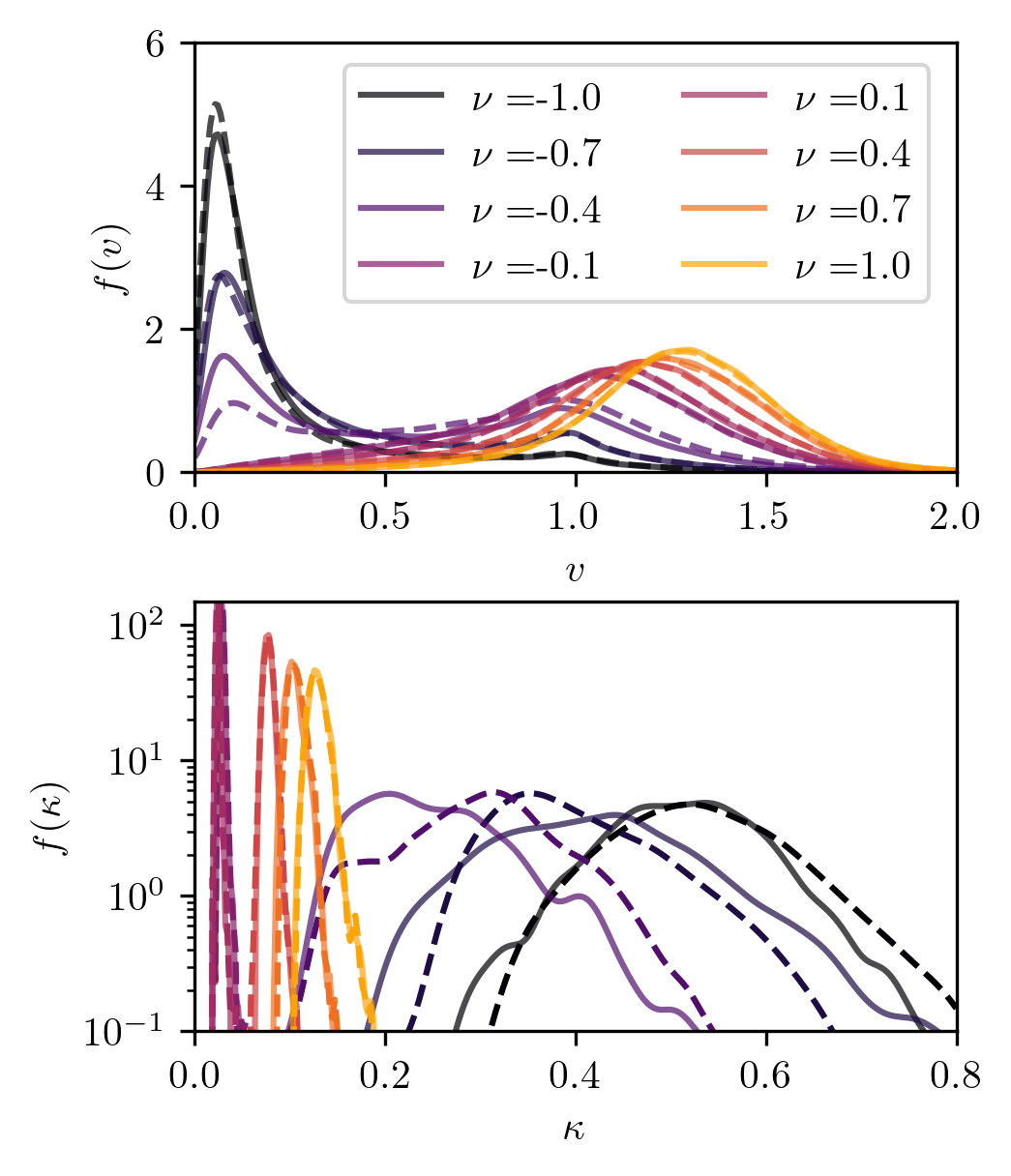}

    \captionsetup[subfigure]{labelformat=empty}
    \begin{picture}(0,0)
    \put(-102,275){\contour{white}{ \textcolor{black}{a)}}}
    \put(-102,145){\contour{white}{ \textcolor{black}{b)}}}
    \end{picture} 
    \begin{subfigure}[b]{0\textwidth}\caption{}\vspace{-10mm}\label{fig:PDFa}\end{subfigure}
    \begin{subfigure}[b]{0\textwidth}\caption{}\vspace{-10mm}\label{fig:PDFb}\end{subfigure}
    \vspace{-5mm}
    
    \caption{PDFs of (a) velocity and (b) rotational activity parameter at various rotational mobility coefficients. The solid line is test data and the dashed line is the HE-GNN model.}
    \label{fig:PDF}
\end{figure}
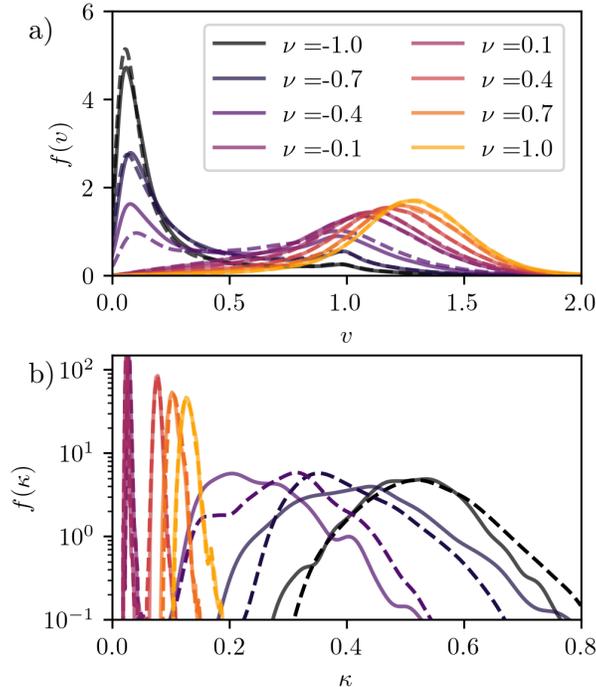

These results show the HE-GNN model qualitatively captures the correct dynamics. We canquantify these results by comparing the microswimmer velocity $v_i=||[dx_i/dt,dy_i/dt]||$ and of the rotational activity parameter $\kappa=(1/N)\sum_{i=1}^N |d\alpha_i/dt|/v_i$ between test and HE-GNN predicted time series.
Figure \ref{fig:PDF} shows the probability density function (PDF) of the microswimmer velocity and of the rotational activity parameter for test data covering 100 time units. We consider these statistics because the aggregation and swirling clearly separate with these PDFs -- which will not be the case for other order parameters we consider. When $\nu$ is negative, the microswimmers aggregate, hindering motion and resulting in low values of $v_i$. When $\nu$ is positive, the microswimmers take on a wider range of $v_i$ values, which is dictated by repeated hydrodynamic and steric interactions. Both the test data and the HE-GNN model give PDFs with averages $v>1$ when $\nu$ is large. The hydrodynamic interactions in the swirling motion assist microswimmers to move faster than without microswimmers (where $v=1$), which the HE-GNN model accurately captures. The PDF of $\kappa$ is more difficult to capture because this quantity is averaged over all microswimmers at a given time. Moreover, this statistic ranges multiple orders of magnitude because $\kappa=0$ when $\nu=0$ (i.e., the microswimmers do not rotate). Again, the $\kappa$ PDF of the HE-GNN model is in good quantitative agreement with the $\kappa$ PDF of the test data. Here, we have shown the HE-GNN model accurately captures the phase transition,

\begin{figure*}
    \centering
    \includegraphics[width=\textwidth]{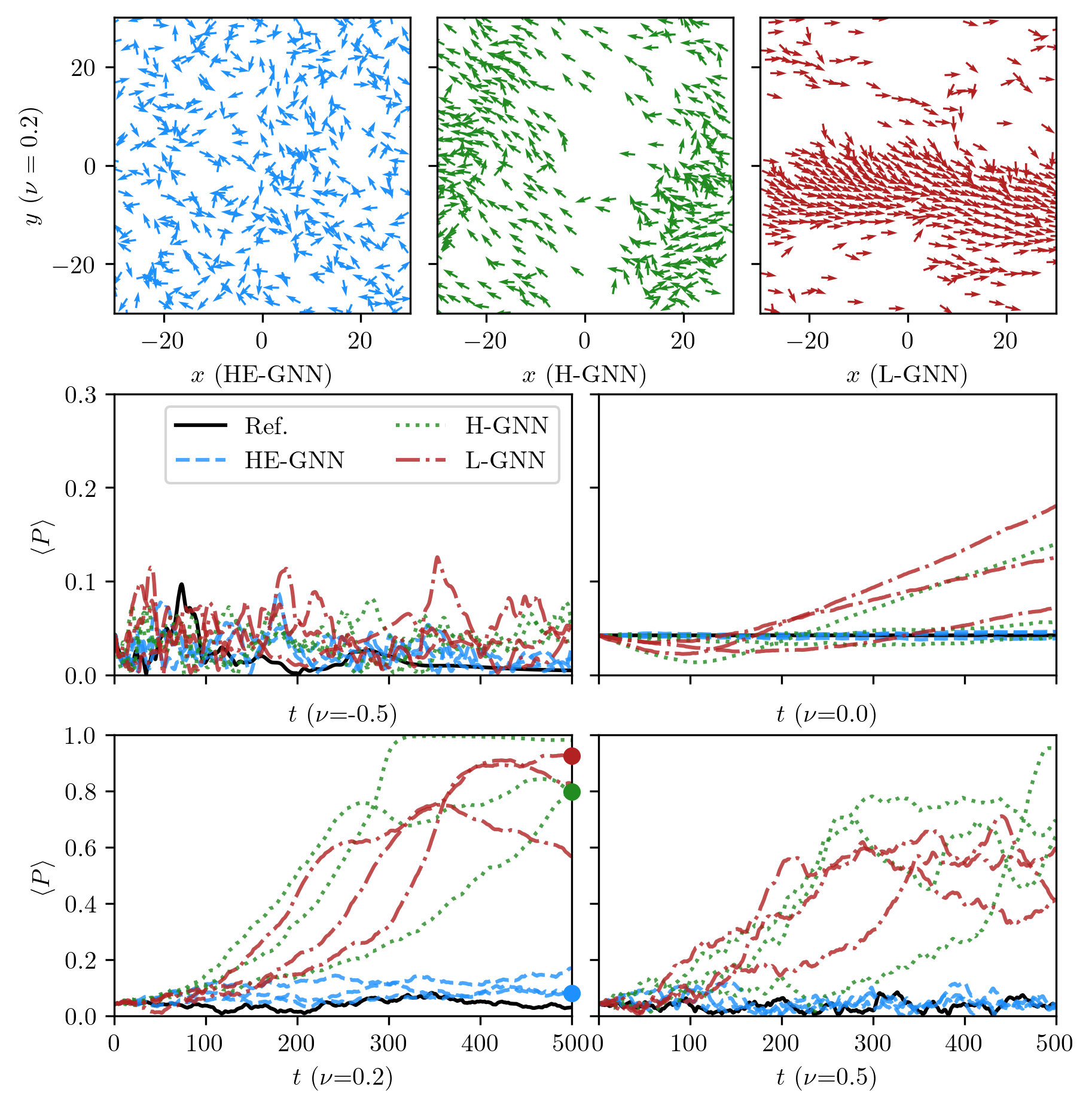}

    \captionsetup[subfigure]{labelformat=empty}
    \begin{picture}(0,0)
    \put(-220,480){\contour{white}{ \textcolor{black}{a)}}}
    \put(-61,480){\contour{white}{ \textcolor{black}{b)}}}
    \put(79,480){\contour{white}{ \textcolor{black}{c)}}}
    \put(-187,319){\contour{white}{ \textcolor{black}{d)}}}
    \put(22,319){\contour{white}{ \textcolor{black}{e)}}}
    \put(-187,172){\contour{white}{ \textcolor{black}{f)}}}
    \put(22,172){\contour{white}{ \textcolor{black}{g)}}}
    \end{picture} 
    
    \begin{subfigure}[b]{0\textwidth}\caption{}\vspace{-10mm}\label{fig:Polara}\end{subfigure}
    \begin{subfigure}[b]{0\textwidth}\caption{}\vspace{-10mm}\label{fig:Polarb}\end{subfigure}
    \begin{subfigure}[b]{0\textwidth}\caption{}\vspace{-10mm}\label{fig:Polarc}\end{subfigure}
    \begin{subfigure}[b]{0\textwidth}\caption{}\vspace{-10mm}\label{fig:Polard}\end{subfigure}
    \begin{subfigure}[b]{0\textwidth}\caption{}\vspace{-10mm}\label{fig:Polare}\end{subfigure}
    \begin{subfigure}[b]{0\textwidth}\caption{}\vspace{-10mm}\label{fig:Polarf}\end{subfigure}
    \begin{subfigure}[b]{0\textwidth}\caption{}\vspace{-10mm}\label{fig:Polarg}\end{subfigure}
    
    \vspace{-5mm}

    \caption{Snapshots of microswimmers predicted using the (a) HE-GNN, (b) H-GNN, and (c) L-GNN at the markers shown in (f). Time series of the polar order parameter are shown for (d)$\nu=-0.5$, (e) $\nu=0$, (f) $\nu=0.2$, and (g) $\nu=0.5$ using the different GNN models. Each model type has three trajectories from three separately trained models.}
    \label{fig:Polar}
\end{figure*}
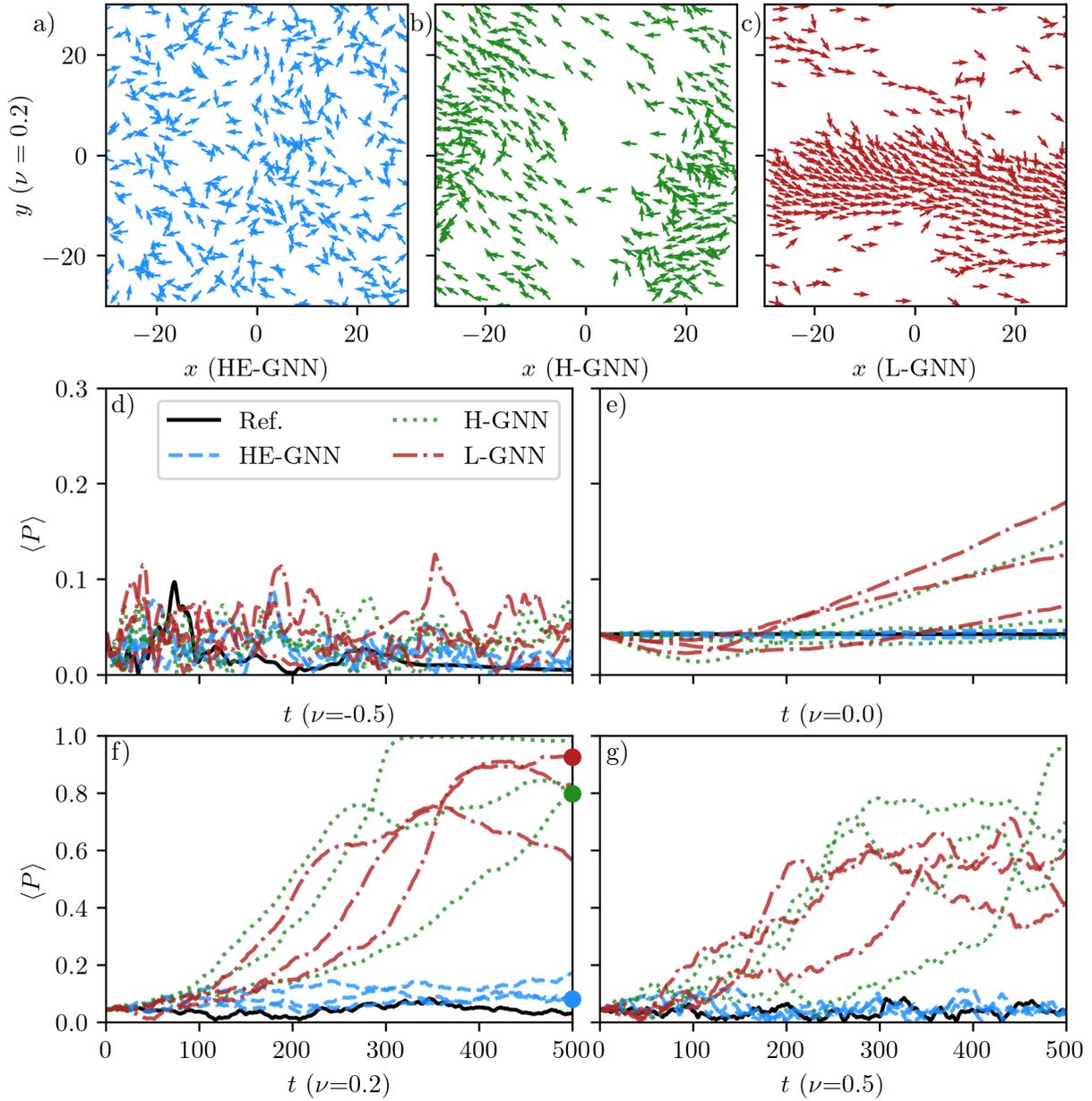

Next, we investigate the importance of hierarchy and equivariance by testing a local GNN and a hierarchical GNN. We set the radius to $R_1=20$ and $R_2=30$. The local GNN (L-GNN) is centered and uses a local graph in radius $R_1$. The hierarchical GNN (H-GNN) differs from the HE-GNN in that it does not account for rotational equivariance. Figure \ref{fig:Polara}-\ref{fig:Polarc} shows snapshots of these three approaches at long-times for $\nu=0.2$. While the HE-GNN model retains the correct swirling dynamics, the H-GNN and L-GNN models cause the microswimmers to align and aggregate.
This alignment can be identified by computing the polar order parameter 
\begin{equation}
    \left<P\right>=\dfrac{1}{N}\left| \sum_{j=1}^Ne^{i\alpha_j(t)}\right|.
\end{equation} 
When $\left<P\right>=1$ swimmers are aligned and when $\left<P\right>=0$ swimmers are oriented randomly. In Fig.\ \ref{fig:Polard}-\ref{fig:Polarg}, we show the time evolution of the polar order parameter for each style of model as $\nu$ varies. We consider this parameter over 500 time units, which is 5 times longer than the training data, to examine if model statistics diverge at long times.

For $\nu<0$ the microswimmers exhibit aggregation in all models, and all models predict low values of $\left< P \right>$. Only the local model exhibits some unexpected peaks in ordering at long times. Adding hierarchical and equivariant information is less important during aggregation. Aggregated microswimmers are largely influenced by the local interactions within a cluster, which makes accounting for long-range interactions less important.
Also, the clusters that do form are circular, making them approximately rotation invariant. Thus, accounting for equivariance is also less important.

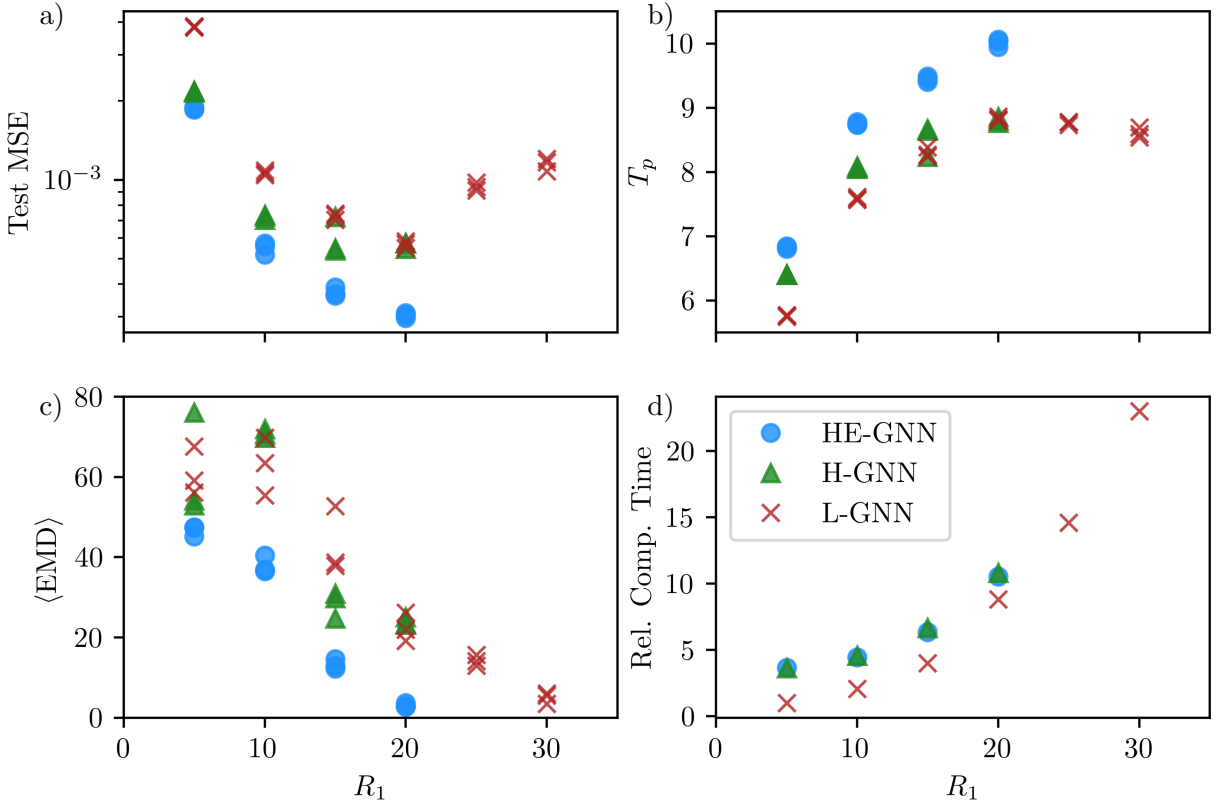
\begin{figure*}
    \centering
    \includegraphics[width=\textwidth]{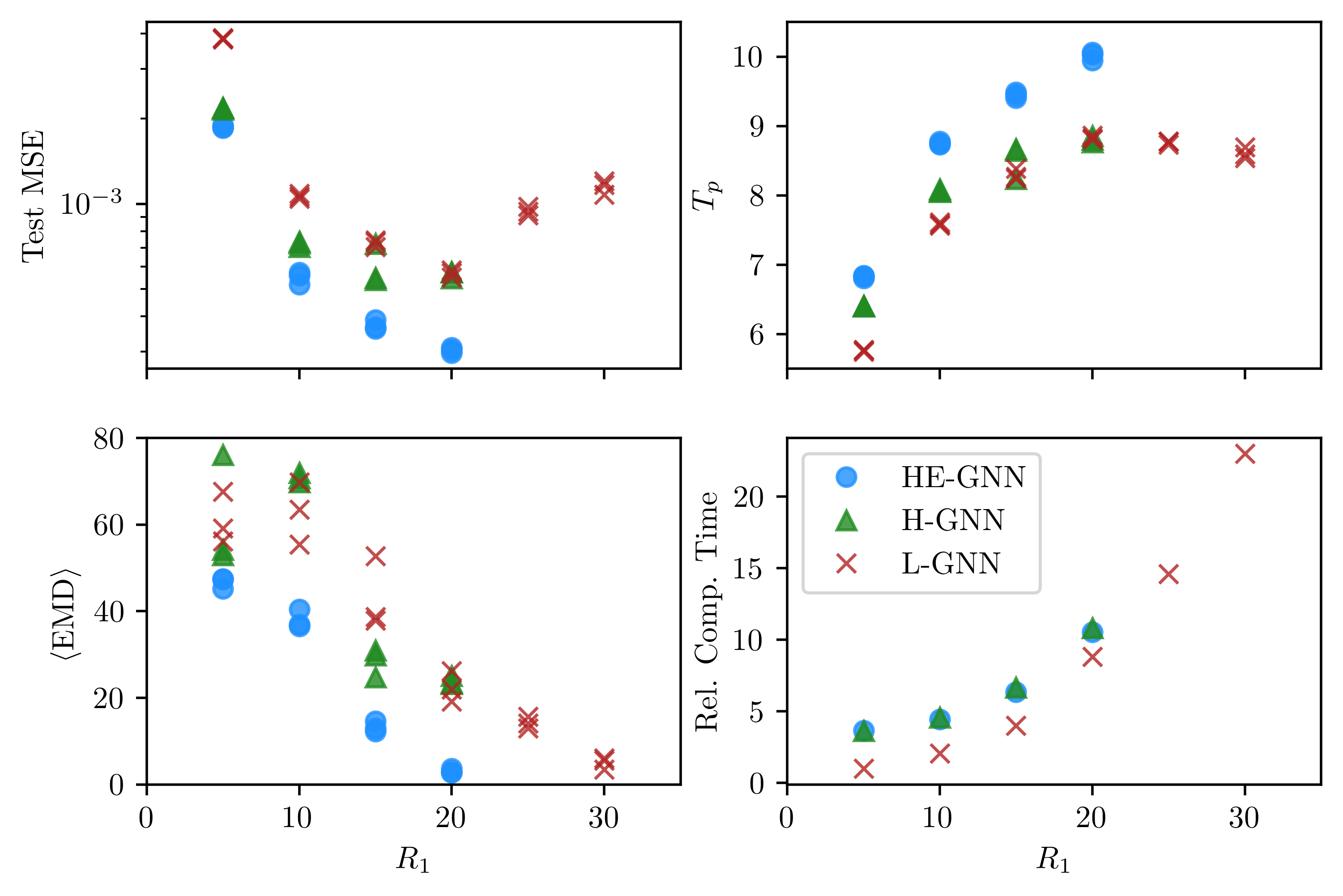}

    \captionsetup[subfigure]{labelformat=empty}
    \begin{picture}(0,0)
    \put(-210,323){\contour{white}{ \textcolor{black}{a)}}}
    \put(20,323){\contour{white}{ \textcolor{black}{b)}}}
    \put(-210,175){\contour{white}{ \textcolor{black}{c)}}}
    \put(20,175){\contour{white}{ \textcolor{black}{d)}}}
    \end{picture} 
    \begin{subfigure}[b]{0\textwidth}\caption{}\vspace{-10mm}\label{fig:Statsa}\end{subfigure}
    \begin{subfigure}[b]{0\textwidth}\caption{}\vspace{-10mm}\label{fig:Statsb}\end{subfigure}
    \begin{subfigure}[b]{0\textwidth}\caption{}\vspace{-10mm}\label{fig:Statsc}\end{subfigure}
    \begin{subfigure}[b]{0\textwidth}\caption{}\vspace{-10mm}\label{fig:Statsd}\end{subfigure}
    \vspace{-5mm}
    
    \caption{Performance of models for varied $R_1$. (a) the test mean-squared error in predicting the ODE, (b) the ensemble-averaged prediction time, (c) the average earth mover's distance between polar order PDFs, and (d) the relative compute time (relative to L-GNN $R_1=5$). Each model type shows results for three separately trained models at each $R_1$.}
    \label{fig:Stats}
\end{figure*}

Significant differences appear between the models for $\nu\geq0$ because long-range interactions influence the swirling motion exhibited by the microswimmers. When $\nu=0$, the microswimmers do not rotate (i.e., the polar order stays constant). All three HE-GNN models maintain a nearly constant polar order, whereas the other models develop a small bias leading to an increase in the polar order at long times. The best hierarchical models maintain a nearly constant polar order, while the local models all diverge substantially. 

For higher values of $\nu$, the deviations from the test data become larger. We show $\nu=0.2$ because the models deviate the most at this value, and we show $\nu=0.5$ for consistency with Fig.\ \ref{fig:Traj}. The results are similar in both the $\nu=0.2$ and $0.5$ cases. Only the HE-GNN model maintains low polar order for all the models (with slightly higher values for $\nu=0.2$). All other models align sometime after 100 time units. The long-time behavior is inconsistent among the other models (i.e., not HE-GNN) because once the alignment increases the trajectory moves outside of states similar to the training data, making the model predictions poor. Notably, the best H-GNN models track longer than the local models, indicating that the hierarchy of graphs can improve results, and all models output reasonable polar order for the first 100 time units, indicating that the strong alignment happens at longer time ranges than seen in the training data.

We end this section by investigating the importance of the radii ($R_1$ and $R_2$) on model performance. Here we fix the radius of the global graph to be $R_2=R_1+10$. In all the comparisons we train three models at each value of $R_1$. In Fig.\ \ref{fig:Statsa}, we show the mean-squared error of predicting $d\mathbf{r}/dt$ (Eq.\ \ref{eq:Loss}). All models tend to improve as the radius increases to $R_1=20$. However, once we increase the radius of the local model further the error increases, likely due to the additional complexity of training with these larger graphs. In most cases, we see a decrease in model error at a given radius when we increase model complexity, indicating that both the hierarchical and equivariant portions of the HE-GNN model play an important role in model performance.

To investigate the effect that this error has on tracking performance, we compute the ensemble-averaged prediction time 
\begin{equation}
    T_p=\text{argmax}_{t_f} \left\{t_f | \left< ||\mathbf{x}'_i(t)-\tilde{\mathbf{x}}'_i(t)||\right><\epsilon, \forall t\leq t_f\right\}.
\end{equation}
This corresponds to when the average distance between swimmers increases past some distance $\epsilon$ for the first time.
Here, we denote the microswimmer position as $\mathbf{x}'_i$ instead of $\mathbf{x}_i$ due to how we account for the periodic boundaries. If we used the microswimmers location between $-L$ and $L$ there could be a large jump in $\mathbf{x}_i(t)-\tilde{\mathbf{x}}_i(t)$ if the swimmer crosses the boundary with the true ODE, but not with the model. To address this, $\mathbf{x}_i'$ corresponds to the position if we did not map the location back across the periodic boundary. Figure \ref{fig:Statsb} shows the ensemble-averaged prediction time for $\epsilon=1$. For this plot, we sample 10 initial conditions (every 50 time units) from the test trajectories for all values of $\nu\geq0$.
We omit $\nu<0$ because the aggregation leads to long tracking times that skew the results. The prediction time increases as we increase the radius and the model complexity. At $R_1=20$, the HE-GNN model performs the best and can track for 10 time units. Once the radius becomes too large, the tracking time starts to decrease for the local model.

The previous two statistics evaluated instantaneous and short-time tracking of the models. To evaluate the long-time predictive capabilities, we compute the earth mover's distance (EMD) \citep{Rubner1998} between the PDFs of the polar order parameter of the test data $F(\left<P\right>)$ and the model $F(\tilde{\left<P\right>})$ and average over all values of $\nu$ for $\left<\text{EMD}\right>$. This concisely evaluates the accuracy of the polar order time series we showed in Fig.\ \ref{fig:Polar}. The EMD is insensitive to binning and small shifts in the PDFs, making it appropriate for this comparison. We compute the EMD by solving a transportation problem in which we find the flow that minimizes the work required to move some ``supply'' to some ``demand''. In this case, the supply corresponds to the true PDF and the demand corresponds to the model PDF, and the work is the flow multiplied by the Euclidean distance between bins in the PDFs. 

Figure \ref{fig:Statsc} shows the EMD for the various models. The HE-GNN models at $R_1=20$ achieve the lowest EMD, which corresponds to the trajectories shown in Fig.\ \ref{fig:Polar}. As $R_1$ decreases the EMD increases for all models, and, again, the trend is typically that increased model complexity leads to better results. At $R>20$ the local models start to perform better despite the worse short-time model performance. This suggests that although a small radius is better for short-time tracking, a large radius is important to capture long-time statistics that may depend upon long-range interactions. The HE-GNN model performs well because it splits these two effects into the local and global graphs, and the equivariance helps by reducing the amount of training data needed for good model performance.

Finally, Fig.\ \ref{fig:Statsd} shows the relative compute time required for each of the models. We compute 20 steps of the ODE solver and normalize by the compute time of the local model with a radius of $R_1=5$ -- the fastest model. When the radius of the local graph is small adding the global graph slows down the computation, however, as we enlarge the radius the computation time increases in a quadratic manner due to the enlarged area of the circle and the main computational cost coming from forward passes of the local graph. For example, if we compare the HE-GNN model at $R_1=20$ (recall $R_2=30$ here) to the local model at $R_1=30$, we see that the local model takes more than twice as long to run, even though both models use information from the same number of swimmers. When comparing these two models, the HE-GNN model has better instantaneous performance, better short-time tracking, and slightly increased accuracy in the polar order parameter while taking less than half the time to run. This highlights the clear benefit of our HE-GNN approach.

\section{Discussion}
\label{sec:Discussion}

Collective dynamics in which groups of agents interact with one another often display long-range, macroscopic properties. Predicting these macroscopic properties can be difficult in part because the underlying dynamics of these systems may be unknown. This makes these systems a natural choice for data-driven modeling. However, as we have shown, care must be taken in constructing these data-driven models, otherwise they fail to capture these macroscopic properties. In particular, we showed the importance of constructing a hierarchy of local and global graphs, and the importance of enforcing equivariance to system symmetries (i.e., rotation and translation). By incorporating these properties into a GNN framework we built accurate data-driven models to predict the dynamics of vortex clusters and a group of microswimmers. 

In the case of vortex clusters, we constructed a local graph connecting all vortices within a cluster, and a global graph connecting all vortices within a cluster to the mean values of neighboring clusters. We then mapped the vortices to a rotation-translation invariant subspace, which the GNN used to predict the velocity of the vortices. These HE-GNNs accurately tracked the vortex centroids and conserved the Hamiltonian for extended periods of time. We compared the HE-GNN model to a fully-connected graph that did not account for equivariance to rotations. In theory, this model should be able to perfectly reconstruct the dynamics. However, when hierarchy and equivariance are not enforced by construction, the model must learn these properties during training. This is a problem for GNN training because the dataset needs to be sufficiently rich in sampling all of the different orientations so that the model predictions do not move far from the training data.

Next, we applied our approach to predict the dynamics of microswimmers. A few key differences between this problem and the vortex dynamics include: the microswimmers are not in clear clusters, the microswimmers have orientation, and the microswimmers undergo a phase transition as the rotational mobility coefficient changes. We constructed a hierarchy of graphs around each microswimmer in which short-range interactions were directly accounted for, and long-range interactions were combined. We then mapped to rotation-translation invariant subspace by centering on a microswimmer and rotating by the orientation of that microswimmer. This graph structure maintained the proper polar order parameter, whereas the removal of the hierarchy or the equivariance caused the microswimmers to spontaneously align. Furthermore, this HE-GNN model was also able to capture the transition from an initially aggregated system to the a swirling system, which never appeared in the training data. Even in this microswimmer simulation, where there does not exist a clear hierarchical structure, separating short- and long-range interactions substantially improves the tracking capabilities of GNN models.

In these systems, the flexibility of GNNs was important. The local and global graphs both contained a variable number of nodes in training and testing. For the swimmers, the node count even changed when considering a single test trajectory. Standard data-driven methods, such as dense neural networks, do not typically handle variable input sizes. Furthermore, treating this problem in terms of a graph was useful because it allowed us to incorporate spatial relationships directly into the graph Laplacian.

\section{Methods}
\label{sec:Methods}

In this section, we provide details on how we perform graph convolutions, account for equivariance, and generate the training for the point vortex and microswimmer problems.

\subsection{Graph Convolutions}
\label{sec:MethodsConv}

One method of performing this graph convolution is with a first-order approximation of a spectral graph convolution \citep{kipf2017} given by
\begin{equation} \label{eq:GConv}
    H^{(1)}=\sigma \left( \tilde{L}X\Theta^{(1)} \right).
\end{equation}
Here, $H^{(1)}=[\mathbf{h}_1,\dots,\mathbf{h}_N]\in \mathbb{R}^{N\times d_h}$, $X=[\mathbf{x}_1,\dots,\mathbf{x}_N]\in \mathbb{R}^{N\times d}$, $\tilde{L}$ is the renormalized Laplacian, $\Theta^{(1)}\in\mathbb{R}^{d\times d_h}$ are the learnable parameters, and $\sigma$ is an elementwise activation function. Our GNNs will consist of repeating $K$ graph convolutions ending in a linear graph convolution (or a standard dense neural network) 
to map $\mathbf{h}^{(K)}_i$ to $d\mathbf{\tilde{r}}_i/dt$. We train the parameters in the filters of these graph convolutions $\Theta^{(i)}$ to minimize Eq.\ \ref{eq:Loss}. For more details on the graph convolution in Eq.\ \ref{eq:GConv} we refer readers to \citet{kipf2017}. We use a slight variation of this method called Chebyshev graph convolutions \citep{Defferrard2016}. Although we select a specific GNN architecture for our problems of interest, the methodology we outline is agnostic to this choice.

\subsection{Determing the Rotation Angle}
\label{sec:MethodsAngle}

Mapping the agents to a rotation-invariant subspace requires a unique phase angle. When this angle is not given by the agent, we can determine this angle using the principal component analysis-based method used in \citet{Xiao2019}. This method involves computing the singular value decomposition of the snapshot matrix
\begin{equation}
    USV^T=[\mathbf{r}_1-\mathbf{\bar{r}}, \dots, \mathbf{r}_N-\mathbf{\bar{r}}]
\end{equation}
and mapping the data to an invariant subspace using the left singular vectors $\mathbf{\hat{r}}_i=U^T(\mathbf{r}_i-\mathbf{\bar{r}})$. Unfortunately, due to the sign ambiguity of singular vectors, this transformation is nonunique. In \citet{Xiao2019}, this problem was addressed by considering every possible sign, concatenating all of the resulting orientations, and using a self-attention module. This type of approach accounts for all rotations and improper rotations (i.e., rotation and reflection), as pointed out in \citep{Li2021}. Orthogonal matrices with a determinant of 1 apply rotation, and orthogonal matrices with a determinant of -1 apply an improper rotation. 

Instead of accounting for all orientations, we directly address this sign ambiguity by modifying the method in \citet{Bro2008}. We also choose to not include improper rotations, although equivariance to reflections may be a desirable property for some problems. The algorithm described in \citet{Bro2008} chooses the sign of the singular vectors, such that they point in the same direction as a majority of the data points. This is achieved by computing the sign of the singular vectors that maximize 
\begin{equation} \label{eq:Sign}
    s_k=\sum_{i=1}^N \text{sign}(U_k^T\mathbf{r}_i)(U_k^T\mathbf{r}_i)^2
\end{equation}
for each singular vector. With this sign, we compute the new left singular vector as $U'_k=\text{sign}(s_k)U_k$. For our purposes, this computation is sufficient to compute a unique left singular matrix that maps all rotations and reflections to the same invariant subspace. The algorithm presented in \citet{Bro2008} also includes steps to compute the sign of the right singular vectors. However, those steps are unnecessary for our objective of disambiguating the sign of the left singular vectors.

Lastly, we avoid improper rotations by multiplying the final singular vector by the determinant ($U'_d=\text{det}(U')U'_d$). If the inclusion of improper rotations is desirable, this step can be skipped. We chose to not account for reflections so that both the vortices and the microswimmers only account for the continuous symmetry. However, in the case of microswimmers in two dimensions, reflections could be accounted for by replacing $U_k$ in Eq.\ \ref{eq:Sign} with the vector normal to the microswimmer direction, and determining the reflection based on the sign of $s_k$.

With this new left singular matrix, we map to the rotation invariant subspace with $\mathbf{\hat{r}}_i=U'^T(\mathbf{r}_i-\mathbf{\bar{r}})$. By performing the rotation with the left singular matrix, we can map to a rotation-invariant subspace for data that lies in $\mathbb{R}^d$ for any $d \geq 2$. In two dimensions, this matrix multiplication is equivalent to performing rotation with the angle given by $\theta_e=\text{atan2}(U'_{1,2},U'_{1,1})$, such that $\mathbf{\hat{r}}_i=R(-\theta_e)(\mathbf{r}_i-\mathbf{\bar{r}})$.

\subsection{Point Vortices}
\label{sec:MethodsVortex}

We generated the point vortex data by numerically solving the Biot-Savart equation
\begin{equation}
    \dfrac{d\mathbf{x}_i}{dt}=\sum_{j=1,j\neq i}^N \dfrac{\gamma_j}{2\pi}\dfrac{\mathbf{\hat{k}} \times (\mathbf{x}_i-\mathbf{x}_j)}{|\mathbf{x}_i-\mathbf{x}_j|^2},
\end{equation}
where $\mathbf{\hat{k}}$ is the out-of-plane unit normal vector. To enforce the conservation of the Hamiltonian, we evolve this ODE forward in time using the explicit extended phase space method described in \citet{Tao2016}. We opted for second-order accuracy (Eq.\ 2 in \citet{Tao2016}) instead of fourth-order accuracy because it conserved the Hamiltonian and reduced the computational cost. Each initial condition was evolved for $T=50$ time units with a step size of $\Delta t=0.001$.

The GNN training data consisted of 200 initial conditions, which we shuffled and performed an 80/20 split for training and validation. Each initial condition consisted of two to five vortex clusters with ten to twenty vortices per cluster. The vortex strength of each vortex was randomly chosen with a mean $\bar{\gamma}=0.1$ and a standard deviation $\sigma_\gamma=0.01$ to match the example considered in \citet{Nair2015}. Each vortex cluster consists of randomly placed vortices within a circle of diameter $D_0=1$. The center of each vortex cluster was randomly placed within a circular region with a diameter of 10$D_0$. We resampled the vortex cluster if it fell within a distance of 2.5$D_0$ from another cluster. This selection was motivated by results in \citet{Eldredge2019}, where he showed that a pair of circular co-rotating patches (clusters) do not mix when separated by a distance of 2$D_0$. We chose to increase this cutoff due to randomly chosen positions and strengths of our point vortices within a cluster. Empirically, we found a separation distance of 2.5$D_0$ did not lead to mixing over the time horizons we considered. 

We performed a sweep of GNN parameters to determine appropriate graph convolutions, layer count, width, and activation functions. The GNN structure that provided the best performance was 5 Chebyshev convolution layers with 3 polynomials \citep{Defferrard2016}. After each layer, we applied a rectified linear unit activation, and then on the final layer we performed a Chebyshev convolution with 1 polynomial (a linear operation). The nodal inputs are $\mathbf{n}_i\in \mathbb{R}^3$, the hidden nodal values are $\mathbf{h}^{(k)}_i\in \mathbb{R}^{64}$, and the outputs are $\mathbf{o}\in \mathbb{R}^2$. The hierarchical and HE-GNN model used the same architecture for both the local and global graphs (with different weights) and the fully-connected model used this architecture for the full graph. We trained these models for 50 epochs using an Adam optimizer with a learning rate of $10^{-3}$ that dropped to $10^{-4}$ at 25 epochs, and a batch size of 50. We used PyTorch Geometric \citep{Fey2019} for training all GNNs. Following training, we tested the model on 50 new initial conditions. We evolved these initial conditions forward $T=200$ time units (four times longer than the training data).

\subsection{Microswimmers}
\label{sec:MethodsSwimmer}

We generated the microswimmer data by considering self-propelled agents that interact with one another through far-field hydrodynamic interactions and near-field steric interactions. The properties $\mathbf{r}_n(t)$ of swimmer $n$ consists of its location $\mathbf{x}_n$ and orientation $\alpha_n$. To model hydrodynamic interactions we treated each microswimmer as a potential dipole that generates the complex flow field
\begin{equation}
\begin{aligned}
w(z)=u_x-iu_y=\frac{\sigma_n e^{i\alpha_n}}{(z-z_n)^2},
\end{aligned}
\label{eq:dipole_flow}
\end{equation}
where $u_x$ and $u_y$ are the velocities in the $x$ and $y$ directions, $\sigma_n$ is the dipole strength, and $z=x+iy$ is the position in the complex plane. These agents are propelled with velocity $U$ and are repelled from each other via near-field Leonard Jones potential $V_n$. We considered microswimmers in a doubly-periodic domain of length $L$ resulting in the equations of motion
\begin{equation}
\begin{aligned}
&\frac{d\bar{z}_n}{dt}=U e^{-i \alpha_n}+\mu \bar{w}\left(z_n\right)+V_n \\
&\frac{d\alpha_n}{dt} = \text{Re} \left[\nu \bar{w} i e^{i\alpha_n}\right],
\end{aligned}
\label{eq:dipole_eom}
\end{equation}
where $\bar{w}=u_x-i u_y$ is the complex velocity, given by
\begin{equation}
\bar{w}=\sum_{n=1}^N \sigma_n \rho\left(z-z_n ; L \right) e^{i \alpha_n}
\label{eq:dipole_db_flow}.
\end{equation}
Here, the function $\rho(z)$ is the Weierstrass elliptic function, defined as
$\rho (z;L) = \frac{1}{z^2}+\sum_{k,l} ( \frac{1}{(z-\Omega_{kl})^2}- \frac{1}{\Omega_{kl}^2})$, with $\Omega_{kl}=kL+lL$, $k,l \in \mathbb{Z} -{0}$.
The translational and rotational mobility of the swimmer are $\mu$ and $\nu$.
We fix $\mu=0.9$, $U=1$, and $\sigma_n=1$, and vary $\nu\in[-1,1]$~\cite{Kanso2015,Tsang2014}. 

The equations of motions \eqref{eq:dipole_eom}, \eqref{eq:dipole_db_flow} are integrated numerically using a fourth-order Runge Kutta integrator. 
The adaptive timestep is chosen to ensure both relative error and absolute error are less than $1 \times 10^{-9}$.
The initial positions of the swimmers are chosen as an equally spaced mesh with random perturbation of half the mesh size. The initial orientations of the swimmers are chosen uniformly randomly from $0$ to $2\pi$. The collective behavior of this model strongly depends on the parameter $\nu$. 
As illustrated in Fig.\ \ref{fig:Traj}, negative $\nu$ leads to aggregation, while positive $\nu$ leads to swirling. 

Here, our training data consists of 21 trajectories of 400 microswimmers all starting from the same initial condition where the swimmers are placed on an equally spaced mesh with random perturbation in space of half mesh size and random orientations. In each trajectory, all microswimmers have the same value of $\nu$ (sample from $[-1,1]$), and we evolve each trajectory forward to $T=100$.  Time is normalized so that a microswimmer without hydrodynamic interactions would move 1 unit length in 1 unit time (i.e., microswimmers driven only by self-propulsion). The domain size is chosen as $60$ unit length. 

From these trajectories, we shuffle and sample $2.5 \times 10^4$ graphs that we split into a 75/25 split for training and validation. Each GNN architecture consists of 5 Chebyshev convolution layers with 3 polynomials and a rectified linear unit activation after each layer. These convolutions map the graph inputs $\mathbf{n}_i\in \mathbb{R}^4$ (local) or $\mathbf{n}_i\in \mathbb{R}^5$ (global) to graphs with hidden nodal values $\mathbf{h}^{(k)}_i\in \mathbb{R}^{64}$. We then sample the node associated with the central microswimmer $\mathbf{h}^{(5)}_0$, append $\nu$ to this vector, and pass this through three dense neural network layers to output $d\mathbf{r}/dt\in \mathbb{R}^3$. Each hidden layer of the dense neural network is in $\mathbb{R}^{64}$, we used rectified linear units and a linear layer on the output. We trained these models for 150 epochs using an Adam optimizer with a learning rate of $10^{-3}$ that dropped to $10^{-4}$ at 70 epochs and $10^{-5}$ at 140 epochs, and a batch size of 50. Following training, we tested the model on a new initial condition for each value of $\nu$ evolved forward for $T=500$ (five times longer than the training data).

\section*{Acknowledgements}

We would like to thank Jeff Eldredge for insightful discussions. This work was supported by the US Department of Defense Vannevar Bush Faculty Fellowship
(grant no. N00014-22-1-2798).

\bibliography{library}

\end{document}